\documentclass[preprint,showpacs,floatfix,preprintnumbers,amsmath,amssymb]{revtex4}

\usepackage{graphicx}% Include figure files
\usepackage{dcolumn}% Align table columns on decimal point
\usepackage{ulem}% pour barrer
\usepackage{color}% pour barrer

\usepackage{bm}% bold math

\begin{document}
\def\twoplus{$2^+$~}
\def\be{\begin{equation}}
\def\ee{\end{equation}}
\def\lb{\langle}
\def\rb{\rangle}
\def\kr{$^{76}$Kr~}
\def\gave{$\langle\gamma\rangle$}

%data set is hfb-5dch.dat 
\def\Ntot{1712~}
\def\Ngood{1628~}
\def\Ngoods{1609~}
\def\neg_corr{84~}
\def\Nspectra{1693~}

\title{Structure of even-even nuclei using a mapped collective Hamiltonian and the 
D1S Gogny interaction}

\author{J.-P.~Delaroche$^{1*}$, M.~Girod$^1$, J.~Libert$^2$,
H.~Goutte$^1$, S. Hilaire$^1$, S.~P\'eru$^1$, N.~Pillet$^1$, and 
G.F.~Bertsch$^3$\\
\,\,\,}

\email{jean-paul.delaroche@cea.fr}
\affiliation{
$^{1}$ CEA, DAM, DIF,
F - 91297 Arpajon, France\\
$^{2}$ Institut de Physique Nucl\'eaire \\
IN2P3-CNRS/Universit\'e Paris-Sud,\\
91406 Orsay Cedex, France\\
$^{3}$ Department of Physics and Institute of Nuclear Theory,
Box 351560 University of Washington Seattle,\\
WA 98915 USA\\}

\date{PRELIMINARY}

\begin{abstract}

A systematic study of low energy nuclear structure at normal deformation is carried
out using the Hartree-Fock-Bogoliubov theory extended by the Generator
Coordinate Method and mapped onto a 5-dimensional collective quadrupole
Hamiltonian.  Results obtained with the Gogny D1S interaction are presented
from dripline to dripline for even-even nuclei with proton numbers 
$Z=10$ to $Z=110$ and neutron numbers N $\leq$ 200.
The properties calculated for the ground states are their charge radii,
2-particle separation energies, correlation energies,
and the intrinsic quadrupole shape parameters. For the
excited spectroscopy, the observables calculated are the excitation
energies and quadrupole as well as monopole transition matrix elements.
We examine in this work the yrast levels up to $J=6$, the
lowest excited $0^+$ states, and the two next yrare $2^+$ states.
The theory is applicable to more than 90\% of the nuclei which
have tabulated measurements. We assess its accuracy by
comparison with experiments on all applicable nuclei where the systematic
tabulations of the data are available.
We find that the predicted radii have an accuracy of 0.6\%,
much better than can be achieved with a smooth phenomenological description.
The correlation energy obtained from the 
collective Hamiltonian gives a significant improvement to the
accuracy of the 2-particle separation energies and to their differences,
the 2-particle gaps.  Many of the properties depend strongly on the 
intrinsic deformation and we find that the theory is especially reliable
for strongly deformed nuclei.  The distribution of values of the collective 
structure indicator $R_{42}={E(4^+_1)}/{E(2^+_1)}$ has a very sharp peak at the value 10/3,
in agreement with the existing data.  On average, the predicted excitation energy and transition
strength of the first $2^+$ excitation are 12 \% and 22\% higher than
experiment, respectively, with variances of the order of 40-50 \%.
The theory gives a good qualitative
account of the range of variation of the excitation energy of the 
first excited $0^+$ state, but the predicted energies are systematically
50 \% high.  The calculated yrare $2^+$ states show a clear 
separation between $\gamma$ and $\beta$ excitations, and the energies
of the $2^+$ $\gamma$-vibrations accord well with experiment.
The character of the $0^+_2$ state is interpreted as shape coexistence
or $\beta$-vibrational excitations on the basis of relative 
quadrupole transition strengths.   Bands are predicted with the properties
of $\beta$-vibrations for many nuclei having 
$R_{42}$ values corresponding to axial rotors, but the shape
coexistence phenomenon is more prevalent.
The data set of the calculated 
properties of \Ntot even-even nuclei, including spectroscopic properties
for \Nspectra of them, are provided in CEA website and  
EPAPS repository with this article \cite{epaps}.
\end{abstract}

% PACS, the Physics and Astronomy
                             % Classification Scheme.
%\keywords{Suggested keywords}%Use showkeys class option if keyword
                              %display desired
\maketitle
%%%%%%%%%%%%%%%%%%%%%%%%%%%%%%%%%%%%%%%%%%%%%%%%%%%%%%%%%%%%%%%%%%%%%%%%%%%    
\section{Introduction}

A present-day goal in nuclear theory is to develop a universal
theory of nuclear structure, in the sense that it is
well-founded in its methodology and is applicable across the
chart of nuclides.  The most promising starting point is
self-consistent mean-field, but the theory must be extended in some way
to treat excitations and nuclear spectroscopy.  For any
candidate theory or methodology, one needs to know its
performance on known observables to be confident about
predictions to unknown nuclei or regions of the nuclear chart. 
It is our purpose to provide and document this information for
one particular theory, the constrained-Hartree-Fock-Bogoliubov
(CHFB) theory together with a mapping to the five-dimensional
collective Hamiltonian (5DCH), using the Gogny D1S interaction
in the nuclear Hamiltonian~\cite{De80,Be91}.  The results presented
here are a major extension of the our study of one particular
observable in the CHFB+5DCH theory, the low-lying quadrupole excitation
\cite{be07}.  Since that work and during the course of the present
calculations
two new parameterizations of the 
Gogny force have been published \cite{hilaire}.
These are aimed at removing systematic deviations of calculated binding energies from 
measurements using the D1S force.  However, for our purposes, the global assessment
of spectroscopy properties, it is more important at this stage to benchmark
the predictions of a stable and widely tested Hamiltonian.  
The present results will therefore 
serve as a baseline for comparisons with the next generation of similar structure calculations 
based on the new parameterizations. We mention that the present era of global calculations within self-consistent mean field
theory using a fixed interaction started with the calculations 
of Tajima et al. \cite{ta96} and
Lalazissis et al. \cite{La96}. 
Also, the mapping to 5DCH has 
recently been adapted and applied to the Skyrme and the relativistic mean-field 
Hamiltonians  \cite{kl08},\cite{ni09}.

The CHFB+5DCH theory is one of several paths that could be taken to
extend mean-field theory to describe spectroscopic properties such as
excitation energies.  Rather than constructing a collective Hamiltonian
from the CHFB solutions, the constrained wave functions may be projected on 
good angular momentum and then used directly
to construct a discrete basis Hamiltonian.  This technology in now fairly
well developed \cite{Bonche,Robledo} and has recently been applied 
to the global study of correlation energies \cite{Ben06} and the
lowest $2^+$ excitations \cite{sa07}.   In principle, it should give
similar results to the CHFB+5DCH theory applying the same Hamiltonian.
However, in the present implementations the discrete-basis wave functions
do not include triaxial
deformations and the effective rotational inertias arising from the
Hamiltonian matrix element are not self-consistent.   On
the other hand, discrete basis methods do not rely on the Gaussian Overlap Approximation
which we require for our Hamiltonian mapping.  There is a
third path to introduce dynamics into
mean field theory, the quasi-particle random-phase
approximation (QRPA).
That has also been applied to many
nuclei, but so far global surveys have been
restricted to spherical nuclei \cite{te08}.

The methodology in the present work has two stages, the CHFB 
calculations to set up the 5DCH input, and the solution of the 5DCH
equations.  For the first part, we perform CHFB calculations 
on a large triaxial grid of quadrupole deformations.  These provide
a potential energy surface and the inertial masses needed
for the 5DCH.  Positive parity solutions are extracted for about 1700
even-even nuclei with proton numbers Z = 10 to Z = 110 and neutron numbers N $\leq$ 200.
These calculations span most of the periodic table from drip-line 
to drip-line.   One purpose of our work is to establish
benchmarks for the accuracy and reliability of the theory for
energies and properties of the low excitations.  We therefore
make systematic comparisons with experimental data,
especially when it is available as a tabulation from a published 
critical review or data repository.
The quantities that we can easily compare are two-nucleon
separation energies and gaps, excitation energies of the 
lowest excited states including yrast spectra up to $J=6$, and
transition rates of the lowest $2^+$ and excited $0^+$ states.
Since no effective charge is involved, our predictions are
free of parameters beyond those contained in the Gogny D1S interaction.  
We hope these calculations will be helpful to
understand the limitations of the theory and ultimately find
improved methodologies and Hamiltonians.  One important
question deals with shell gaps and magic numbers.  Far from stability, dedicated
experiments have shown that the N = 20, 28 shell gaps experience
erosion and that N = 16 may become magic number at the oxygen
neutron drip-line \cite{st04}. Other experiments are underway to investigate
whether shell quenching takes place too in the vicinity of the N
= 50 gap, a critical issue for the path followed by the
r-process of the nucleosynthesis.  The predictions for the shell
gaps and associated observables coming from the Gogny interaction 
have previously been reported  \cite{Pe00,Ob05}. 

Another purpose of this work is to provide a set of predictions 
for nuclei to be studied in future.
The advent of unstable nuclear beam facilities 
has opened up a new and exciting
area in exploring the structure of exotic nuclei.  Of particular
interest are the nuclei near or at the border lines of stability,
the proton and neutron drip-lines. These nuclei often display
properties which are not present in nuclei located  in the
vicinity of the $\beta$-stability line, and questions are raised
as to whether the nuclear structure models and effective forces tailored over the
past seventy years remain valid in the present context. 
Thus, strong deviations of the experimental findings with
respect to the benchmarked predicted accuracy would signal 
new phenomena in nuclear structure.

Several caveats should be mentioned that limit the domain of
validity of the theory. A basic approximation made
here is to require the HFB fields to conserve parity and
signature.  There is no strong evidence that these symmetries
are violated in the HFB ground states, but there may be some
nuclei for which it happens.  Another limitation arises from the
neglect of two- and higher-order- quasiparticle (qp) excitations. 
The GCM theory (with or without the Gaussian overlap approximation) does not include 
these degrees of freedom which
will inevitably affect the spectrum at higher excitation
energies.   Also, the Hamiltonian is an 
adiabatic one, with parameters calculated in the vicinity of
zero rotational frequency.  This affects the reliability of the
calculated excitations with high angular momentum. 
Finally, the application of the  5DCH requires
further that the overlaps be semilocalized in the
$(\beta,\gamma)$ quadrupole deformation plane (coherence length be small compared to
dimensions of the arena in $(\beta,\gamma)$ in which the wave
function has significant amplitude). Indeed, the mapping of the HFB
to the collective Hamiltonian is problematic for very rigid nuclei, such as the
doubly magic ones. For these reasons, we concentrate on the 
lowest excited states in this work in non-doubly magic nuclei, and
restricting angular momentum to $J\le 6$.

\section{Reminder of Formalism and Computational Implementation}

For completeness, we recall here the equations to be solved.
The derivation and some aspects of the implementation are
presented in more detail in Ref. \cite{Li99}.
The potential energy surface that goes into the 5DCH is obtained from
constrained Hartree-Fock-Bogoliubov calculations (CHFB) based on the 
Gogny D1S interaction.
The CHFB equations to solve are
\begin{equation}
\label{eq:chfb}
\delta \langle \Phi(q_{0},q_{2})|\hat{H}-\lambda_0 \hat{Q}_0 -\lambda_2 \hat{Q}_2 -\lambda_Z \hat{Z}-\lambda_N \hat{N}
|\Phi(q_{0},q_{2})\rb = 0.
\end{equation}
$\hat{H}$ is the Hamiltonian, and the other terms are 
linear constraints to obtain particle numbers $N,Z$ and quadrupole 
moments $q_0,q_2$ according to 
\begin{eqnarray}
\langle \Phi(q_{0},q_{2})| \hat{Q}_i |\Phi(q_{0},q_{2})\rb = q_i,\\
\nonumber
\langle \Phi(q_{0},q_{2})|\hat{Z}(\hat{N}) |\Phi(q_{0},q_{2})\rb = Z(N).
\end{eqnarray}
Here we define the quadrupole operators as $\hat{Q_0}=2z^2-x^2-y^2$ and
$\hat{Q_2}=x^2-y^2$. 
The CHFB equations are solved for each set of deformations by expanding the 
single particle states in a 
triaxial harmonic oscillator (HO) basis. 

The triaxial oscillator basis is subject to truncation according to 
  $$(n_x+1/2) \hbar\omega_x +(n_y+1/2) \hbar\omega_y+(n_z+1/2) \hbar\omega_z \le (N_0+2) \hbar\omega_0,$$
where $(\hbar\omega_0)^3= \hbar\omega_x\hbar\omega_y\hbar\omega_z$, 
with $\hbar\omega_i (i = x, y, z)$ as oscillator  basis parameters, and $n_i (i = x, y, z)$ as quantum numbers.
The basis is determined as follows. For a nucleus with Z protons and N neutrons, the number $N_0$ is such that
 $\mathcal{N}$, the number of single particle states in the Hartree-Fock scheme, 
 is eight times the number of levels occupied by the larger among the Z or N values. $\cal{N}$ is a function of $N_0$, 
 fulfilling the empirically established equation 
$$ \mathcal{N}= 2.1  N_0^2 + 0.0072 N_0^4,$$
from which $N_0$ is deduced. In general $N_0$ is not an integer.

Instead of using the $\hbar\omega_i$ parameters, we adopt in the HFB calculations the parameters $\hbar\omega_0$, P and Q, 
with P= $\hbar\omega_x/\hbar\omega_y$   and Q=$\hbar\omega_x/\hbar\omega_z$. 
These parameters need be determined to define the oscillator basis at each point of the grid. 
The parameters P and Q are determined using formulas based on a liquid drop parametrization 
of nuclear shape. These formulas depend upon $\beta$ and $\gamma$ deformations in the constrained HFB calculations, and write as
\begin{eqnarray}
\label{pq}
P=\exp[-x\sqrt 3\sin\gamma],\\ \nonumber 
Q=\exp(x[\frac{3}{2}\cos\gamma-\frac{\sqrt 3}{2}\sin\gamma]), 
\end{eqnarray}
where $x=\beta/(2\beta+1)$.  

$\hbar\omega_0$ is obtained through minimization of the HFB energy. 
This is made for $\gamma$= 0$^o$ and  60$^o$ at a fixed $\beta$ to take advantage of our axially symmetric HFB code which is running much faster than the triaxial code. 
To get $\hbar\omega_0$ values over the triaxial plane, 
we use the interpolation formula 
$$ \omega_0(\beta, \gamma)=\frac{1}{2}[\omega_0(\beta,\gamma=0^o) + 
\omega_0(\beta,\gamma=60^o)]+
\frac{1}{2}[ \omega_0(\beta,\gamma=0^o) - \omega_0(\beta,\gamma=60^o)] \cos(3\gamma).$$

For the basis truncation, a fine tuning of P and Q values is performed so as to maximize the number of particle states without 
altering the numbers of oscillator shells in each of the three directions as obtained with Eq.(\ref{pq}).
Typically, the number of major shells $\cal{N}$ ranges from 6 to 16 in the present study.

The Bogoliubov space is restricted by imposing the self-consistent symmetry $\hat{T}\pi_2$, with $\pi_2$ the reflection with 
respect to the x0z plane, and $\hat{T}$ the time-reversal symmetry. The HFB nuclear states have also been taken invariant
under the left-right symmetry ~\cite{Gi83}.
One technical point should be mentioned.  Since there are many points
to calculate, it is important to have an efficient algorithm to perform
iterative solution of the CHFB equations.   From the early days, we found it very helpful 
in this respect to use first order perturbation theory to update the
linear constraints. During the
iterative procedure the obtained mean value $q_j$ differs from the imposed value $q^{(0)}_j$, the corrections 
applied to the Lagrange parameters are~\cite{Be85}
\begin{equation}
\delta \lambda_i = \sum_{j=0,2} ( {\cal M}_{-1}^{ij})^{-1} (q_j^{(0)}-q_j). 
\end{equation}
The moments $\cal M$ of the off-diagonal quadrupole operators in
the constrained HFB configurations are defined as
\begin{equation}
{\cal M}^{ij}_k(q) = \sum_{\mu\nu} \frac{\lb \Phi_q | \eta_\mu\eta_\nu \hat Q_i |
\Phi_q\rb \lb \Phi_q | \eta_\mu\eta_\nu \hat Q_j | \Phi_q\rb}{(E_\nu + E_\mu)^k },
\label{moment}
\end{equation}
where $\mu,\nu$ label quasiparticles with destruction operators $\eta$ and
energies $E_{\mu}$ and $E_{\nu}$, respectively.

The potential energy surface is then determined from the expectation
value of the Hamiltonian, corrected for the one- and two-body center-of-mass energy
\be
V(q_0,q_2) = \langle \Phi(q_{0},q_{2})|\hat{H}-\frac{\hat P^2}{2 m A}|\Phi(q_{0},q_{2})\rb.
\label{pot}
\ee
It is convenient to use the dimensionless deformation parameters $(\beta,\gamma)$ which
are defined through $\displaystyle \beta = \sqrt{5\pi}\frac{\sqrt{q_0^2 + 3 q_2^2}}{3A^{5/3}r_0^2}$ 
and $ \displaystyle \gamma= \arctan \sqrt{3} \frac{q_2}{q_0}$, with $r_0=1.2$~ fm.
Typically, the constrained  HFB equations are solved
on the domain ($0 <\beta < 0.9$ ; $0 < \gamma < \pi/3$) with mesh 
spacings $\Delta \beta = 0.05$ and $\Delta \gamma = 10^\circ$.

The final 5DCH is expressed  \cite{Li99}
\begin{equation}
\hat{H}_{coll}=\frac{1}{2}\sum_{k=1}^3\frac{\hat{J}^2_k}{\mathcal{J}_k}-\frac{1}{2}
\sum_{m,n =0 \, and  \, 2} D^{-1/2}\frac{\partial}{\partial a_m}D^{1/2}(B_{mn})^{-1} \frac{\partial}{\partial a_n}
+V(a_{0},a_{2})-\Delta V(a_{0},a_{2}) ,
\label{eq1}
\end{equation}
where we have made another change of deformation parameters from $(\beta,
\gamma)$
to $a_0= \beta \cos \gamma$ and $a_2 = \beta \sin \gamma$, and where $D$ is the metric
\cite{Ku67}.
There are 3 rotational inertia and 3 quadrupole mass parameters in the
5DCH.  These are all computed from the local properties of the CHFB
solutions at the grid points.  To calculate the rotational inertia 
we implement additional constraining fields $\omega \hat J_k$ to Eq. (1),
where $\hat J _k$ is the angular momentum operator about the $k$ axis.
Calling the new self-consistent solution $\Phi^\omega_q$, we calculate
the inertias $\mathcal{J}_k$ as 
\be
\mathcal{J}_k = \frac{\lb \Phi^\omega_q |\hat{J}_k | \Phi^\omega_q \rb }{\omega}.
\ee
In the limit $\omega \rightarrow 0$, this expression is equivalent to
the Thouless-Valatin inertia.  In practice we take $\omega = 0.002$
MeV to approximate the limit.
The quadrupole mass parameters $B_{ij}$ are calculated in the cranking 
approximation \cite{Li99},
\be
B_{ij}(q)= \frac{1}{2}\frac{{\cal M}^{ij}_{-3} (q)}{( {\cal M}^{ij}_{-1}(q))^2},
\ee
with ${\cal M}^{ij}_k(q)$ the moment defined in Eq.(\ref{moment}). \\
It is important to mention that the cranking approximation
is not self-consistent in the sense that the dynamical rearrangement is not taken into account 
and we should expect some deficiencies in the
theory as a result.

The zero-point energy (ZPE) correction to the potential, Eq.(\ref{pot}), is associated with the 
nonlocality in the quadrupole coordinates. It is calculated according to the
formulas given in Refs. \cite{Gi79,Li99}, 
$$
\Delta V(q) = \frac{1}{4} \sum_{i,j} \frac{ {\cal M}^{ij}_{-2}(q)}{{\cal M}^{ij}_{-3}(q) }.
$$
Here the sum runs over the sets $(i,j)=(0,0),(2,2),(0,2)$ for the vibrational
ZPE and $(i,j)=(1,1),(-1,-1),(-2,-2)$ for the rotational ZPE, 
following the notation of \cite{Li99}.  This includes only the part of
the ZPE arising from the kinetic energy operator.  There is also
a part due to the potential, which we neglect.  This is expected
to be small in typical situations with shallow minima in the 
potential energy surface; it might be significant near magic 
numbers where the curvature of the surface is higher.

Eigenstates and eigenenergies are
obtained as numerical solutions of 
\begin{equation}
\hat{H}_{coll}|JM\rb=E(J)|JM\rb.
\end{equation}
The orthonormalized eigenstates $|JM\rb$ with angular momentum J
and projections M on the third axis in the laboratory frame are expanded as
\begin{equation}
|JM \rb=\sum_K g_K^J(a_0,a_2)|JMK\rb,
\end{equation} 
with $|JMK\rb$ a
superposition of Wigner rotation matrices.  The probability $P(K)$ of
the different $K$ components of the wave function gives a useful
indicator of its character.  This is defined as
\be
P(K)= \int d a_0 d a_2 |g^J_K(a_0,a_2)|^2.
\label{eq:P(K)}
\ee
We refer to Ref. \cite{Li99} for further numerical details on solving
the 5DCH equations. Here we use the value $m_{max}$ = 28 for the order parameter in the power expansion of vibrational amplitude.
This secure a 2\% precision on relative energies in collective spectra.
We calculate radii and quadrupole matrix elements
assuming that the coordinate operators are local in the collective
coordinates \cite{Ku67}. For example, the matrix element 
of the quadrupole
operator ${\cal M}_m = \sum_i^Z r_i^2 Y^2_m(\hat r)$ is calculated as
\be
\langle J' M'  |{\cal M}_m | J M \rangle=\left((2 J +1)(2 J' +1)\right)^{1/2}
\times
\ee
\be
\sum_{K,K',k} \left(\begin{matrix} J'& 2 & J \cr -M'& m&M \end{matrix}\right)
\left(\begin{matrix} J'& 2& J \cr -K'&k& K \end{matrix}\right)\int d a_0 d a_2 g^{J'}_{K'}(a_0,a_2)g^{J}_K(a_0,a_2)
\langle\Phi(a_0,a_2)
| {\cal M}_{k}| \Phi'(a_0,a_2)\rangle .
\ee
This is an approximation, but we have no reason to
doubt its reasonableness.

The correlation energy is defined as
\begin{equation}
E_{corr}= E^{min}_{HFB}-E_{5DCH},
\label{eq:ecorr}
%  correlation energy is positive.
\end{equation} 
where $E^{min}_{HFB}$ is the minimum of the energy at the HFB level, and 
$E_{5DCH}$ is the energy of the collective ground state obtained from the
5DCH calculations.   For nuclei near magic numbers, the calculated
correlation energy may come out negative, which is unphysical.  We have 
kept these nuclei in the accompanying table, but we exclude them when 
we compare the calculated properties with experiment.

Also, the accuracy of the calculations will not be
as high at the extremes of the nuclear chart, due to the incipient
shape instability associated with fission, as well as the limitations
of the harmonic oscillator basis for dripline orbitals.  In the
accompanying tables, we include the ground state properties when
the calculated deformation is consistent with a non vanishing 
fission barrier in the $(\beta,\gamma)$ plane.  This will include some
nuclei that would have vanishing fission barrier when more shape 
degrees of freedom are permitted.

\section{Examples}
To show the scope of the theory, we begin with two examples of nuclei that 
illustrate the complexity of nuclear structure that can be addressed
with the 5DCH.
The first is \kr, which is considered as an example of a soft
nucleus. The second is
$^{152}$Sm, which has a near rotational spectrum but is also considered to
be a transitional nucleus.
\subsection{\kr}
We begin with \kr, a nucleus with a complex spectrum of low-lying
excitations providing evidence for shape coexistence phenomena.
%76Kr:  N=40, Z=36
A few calculated spectroscopic properties of this nucleus were
already reported in Ref. \cite{cl07}.
The ground state of \kr is spherical in the HFB approximation, but becomes 
highly deformed
in the CHFB+5DCH wave function, with mean deformation values of
$\langle\beta\rangle  = 0.33$ and $\langle \gamma \rangle= 24^\circ$. 
The variances of the deformations  are also of interest, namely
\be
\label{eq:db}
\delta \beta = \sqrt{\lb \beta^2\rb - \lb \beta \rb^2};\,\,\,
\delta \gamma = \sqrt{\lb \gamma^2\rb - \lb \gamma \rb^2},
\ee
where $<\gamma^2>$ and $<\gamma>$ are calculated over the sextant $0 <\gamma < \pi/3$.
The values of these quantities in the \kr~ground state are
$\delta \beta = 0.10$ and $\delta \gamma =13^\circ$, suggesting that
the nucleus is fairly rigid in $\beta$ but with some soft 
triaxiality.  Due to the triaxiality, one does not expect to
see a rigid rotor spectrum, despite the large deformation.

\begin{figure*}
\includegraphics [width = 16cm]{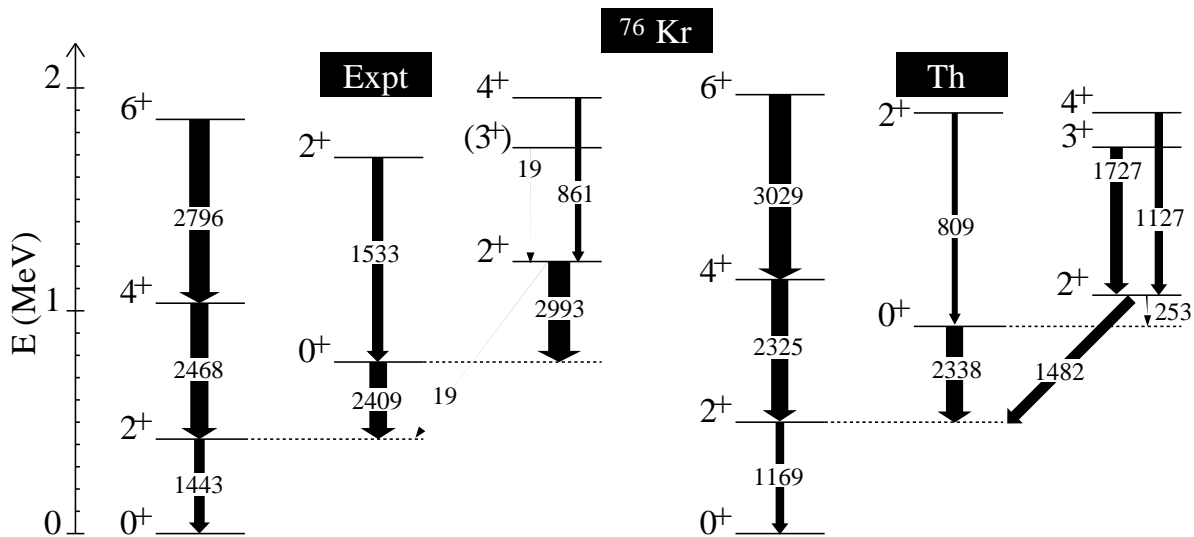}
\caption{ Experimental and theoretical spectra (MeV) and transition strengths ($e^2 fm^4$) 
of $^{76}$Kr showing the excitations
that we examine in the present global study.  The experimental spectrum,
on the left, is from Ref. \cite{cl07} as well as from data repository for the
$3^+$ and $4^+$ members of the $\gamma$ band \cite{brookhaven}. Calculated values are those from Ref. \cite{cl07}.
}
\label{fi:76kr}
\end{figure*}

In this work, we will examine systematically the  
$0^+_1, 2^+_1,0^+_2,2^+_2,4^+_1,   2^+_3,3^+_1 $,  and
$6^+_1$ excitations.  These are shown for the \kr~nucleus together
with the additional states that could form a $\gamma$-vibrational
structure in  Fig. \ref{fi:76kr} \cite{cl07,brookhaven}.
The experimental spectrum is also shown in the figure, and one sees that
excitation energies are reproduced very well. 
The calculated excitation energy of the first excited state, the $2^+_1$, is 
only 21\% higher than experiment.  The other states are proportionally
even closer: the energies of the yrast $4^+_1$ and $6^+_1$ states
are within 10\% of the experimental values.  Even the non-yrast excited
state energies come out well.  We shall later examine systematically the 
$0^+_2,2^+_2$ and $2^+_3$  excitations; in \kr~their predicted energies
are all within 20\% of experiment.  
For the transition
strengths, the predicted  $B(E2;2^+_1\rightarrow 0^+_1)$ is within 20\%
of the experimental value and the higher transitions
along the yrast ladder are within 10\%.

The second excited state in the \kr system is the $0^+_2$ level at 0.77 MeV.
The calculated energy is 0.92 MeV, close enough to make a correspondence
between the two states.  Its mean deformation parameters are close
to those of the ground state, suggesting a $\beta$-vibrational 
interpretation.  The transition rate to the $2^+_1$ state is large
and in very good agreement with experiment.
The $2^+_3$ excitation corresponds in excitation energy
fairly well to experimentally measured state.  The $2^+_3$ wave function
has a large probability $P(K=0)$, suggesting that it be placed with
the $0^+_2$ level as member of the K = 0 excited structure.  Its transition strength
to the $0^+_2$ is large and in qualitative accord with experiment.
Experiment and calculation for the spectroscopic quadrupole moment Q$(2^+_3)$ are
also in accord for both magnitude and sign. The sign is opposite to that for Q$(2^+_1)$, 
nullifying any interpretation  of the K = 0 excited structure as $\beta$-vibrational band and
giving  weight to the interpretation of shape coexistence between prolate and oblate band structures.

From the energetics, the $2^+_2$ level might be assigned at a two-phonon
excitation of the ground state.  The calculated $2^+_2$ wave function has a large probability for $K=2$
($P(2)=0.77$),
suggesting that this level instead is the bandhead of a $\gamma$-structure. 
However, the experimental data on the transition strengths between
the $2^+_2$ and the $0^+_2,2^+_1$ and $3^+_1$ states are very far from the
theoretical predictions. Since transition strengths of $\gamma$-vibrations are very
sensitive to $K$-band mixing, the disagreement of transition strengths
does not rule out the $\gamma$-vibrational interpretation. For more discussions, see \cite{cl07}.

We finally mention the highest excitations, some of which will be beyond
the scope of our global survey.
There is good accord between theory and
experiment for the energetics of the $6^+_1$ state as well as for those for the $3^+_1$ and $4^+_2$ levels
which both form a quasi-$\gamma$ band structure on top of the $2^+_2$ state.  The $6^+_1$
 and $4^+_2$ states have strong transitions to the $4^+_1$ and $2^+_2$ states, respectively, in both
theory and experiment.  

To summarize, the CHFB+5DCH theory provides a very good description of 
low-lying excited states 
\kr spectrum.  While not all aspects are reproduced, many of the energies
and relative transition strengths are given to good accuracy.  The
complex spectra of the Kr isotopes have often been discussed as a 
shape coexistence phenomenon, and the theory does rather well in describing
 these features as well as shape transitions in this region \cite{gi09}.

\subsection{$^{152}$Sm}
\begin{figure}
\includegraphics [ height= 8cm, angle=-90]{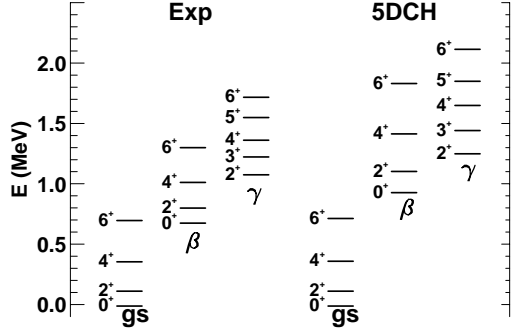}
\caption{\label{fi:sm152} Experimental and theoretical spectra (MeV) of $^{152}$Sm.
The experimental spectrum is based on Ref. \cite{brookhaven} .
}
\end{figure}

The nucleus $^{152}$Sm lies at the start of the deformed lanthanide
region of the nuclear chart and is considered as landmark in the 
identification of first-order quantum phase transition between
spherical and axially deformed nuclei \cite{ia98,li09}.  Its experimental level scheme is shown
in the left-hand panel of Fig. \ref{fi:sm152}, taken from Ref. \cite{brookhaven} and our
calculated level scheme is in the right-hand panel. We first note that the yrast band is well 
reproduced. The $2^+_1$ excitation energy is within 2\% of the experimental one,  and the experimental ratio of the $4^+_1$ to the
$2^+_1$ energies is $R_{42}=3.0$, slightly lower than the rigid axial rotor value 10/3. The 5DCH ratio is 3.0, reproducing
the slight deviation from rigidity.   The yrast spectrum is intermediate between that for harmonic vibrators
and axial rotors, consistent with predictions from the X(5) model designed as analytic description of critical point structures
in N $\simeq $ 90 isotones \cite{ia01,ca01}. For a review see \cite{ca07}.

\begin{table}
\caption{Experimental and theoretical E2 transition strengths $B(E2;J_i\rightarrow J_f)$\ ($e^2fm^4$) of $^{152}$Sm. 
From left to right : exp1 and exp2 are for $B(E2)$ experimental data from Ref.\cite{ku07} and \cite{za99,kl00,ca01,bi03,ca07}, respectively, and th are for 5DCH calculations.}
\begin{tabular}{ cccccc}
\hline\hline
$J^\pi_i$& $\rightarrow $&$ J^\pi_f$&
exp1&
exp2&
th\\
\hline
$6^+_1$  &&  $4^+_1$ &11795(347) &11805(241) &12042  \\
$4^+_1$  &&  $2^+_1$ & 10061(277)  & 10071(144)   &  10171   \\
$2^+_1$  &&  $0^+_1$ &  6938(144) & 6938(144)   & 6671    \\
\hline
$0^+_2$  &&  $2^+_1$ & 1589(194)  & 1590(110)   & 2539    \\
\hline
$2^+_2$  &&  $0^+_2$ & 8048(763)  &  5156(1300)  &   4732  \\
         &&  $4^+_1$ & 867(97)  & 915(96)   &  768   \\
         &&  $2^+_1$ & 277(28)  &  265(24)  &   1066  \\
         &&  $0^+_1$ &  46(4) &  43(5)  &   1.2  \\
\hline
$4^+_2$  &&  $2^+_2$ & 12488(2081)  &  9830(1831)  &  7505   \\
         &&  $6^+_1$ &  763(208) & 193(96)   &  655   \\
         &&  $4^+_1$ & 291(55)  & 260(63)   & 977    \\
         &&  $2^+_1$ & 36(6)  &  48(9)  &   27  \\
\hline
$2^+_3$  &&  $0^+_1$ & 179(12)  & 174(8)   &   430  \\
         &&  $2^+_1$ & 451(28)  &  448(24)  &  77   \\
         &&  $4^+_1$ & 34(3)  & 38(2)   &  686   \\
         &&  $0^+_2$ & 2(0.2)  & $<$2.4   &  1376   \\
         &&  $2^+_2$ & 604(42)  & 1301(193)   & 7382    \\
\hline	   
$3^+_1$  &&  $2^+_1$ &   &  337-819  &   479  \\
         &&  $4^+_1$ &   &  337-867  &   389  \\
         &&  $2^+_2$ &   &   $<$25 &    1974 \\
         &&  $2^+_3$ &   &  2987-38451  &  6627   \\
\hline	   
$4^+_3$  &&  $2^+_1$ &   & 28(8)   &  228   \\
         &&  $4^+_1$ &   & 265(77)   &  4384   \\
         &&  $6^+_1$ &   &  58(19)  &   663  \\
	   &&  $2^+_2$ &   &  9(3)  &   140  \\	
         &&  $4^+_2$ &   &  $<$1686  &  1209   \\
         &&  $2^+_3$ &   &  2409(723)  &   5045  \\
         &&  $3^+_1$ &   &  $<$12046  &   4384  \\	   	   	      
\hline\hline

\end{tabular}\label{tabD}
%\end{ruledtabular}
\end{table}

We now come to the predictions for the $0^+_2$ excitation and the collective structure built upon it.
   The calculated deformation of that
state is $\langle\beta\rangle= 0.29 $, almost the same as the ground
state deformation, $\langle\beta\rangle= 0.30$.  This suggests an 
interpretation as a $\beta$-vibration.  One expects that the
fluctuation in $\beta$ would be larger in the vibrational excitation
than in the ground state; in the harmonic limit, 
$<n|\beta^2|n> = (n+1/2)/B_{00}\,\omega$, giving a ratio of $\sqrt{3}$. 
In fact, the fluctuation in the calculated
wave functions is larger for the excited state, by a factor 1.58.   
Thus, the theoretical wave function has the main characteristics to
be a $\beta$-vibration.

Comparing experimental and calculated spectra, we first note that the calculated 
excitation energy of the $0^+_2$ state is quite a bit higher than observed
experimentally, by nearly 40\%. As we will see later, this is a common 
feature of the CHFB+5DCH theory as presently implemented. 
The experimental energy splitting between the $2^+_2$ and $0^+_2$ states is nearly identical to
that of the ground state band, 
while the theoretical splitting
is larger by 40\%.   This discrepancy is in keeping with that of the X(5) model \cite{ca07}.
We conclude that the theory confirms in an approximate
manner the existence of a band structure based on the $0^+_2$ excitation,
but in detail deviates from the $\beta$-vibrational limit in
the in-band energetics.

The CHFB+5DCH theory also predicts a $2^+_3$ excitation and collective structure upon it. This sequence is interpreted as a quasi $\gamma$-vibrational band, with head level energy slightly higher than that observed in this nucleus. Our calculated third $2^+$ state has the same average $\langle \beta \rangle$ deformation as the ground state,
supporting a vibrational interpretation.  If it were a true $\gamma$-vibration,
it should have high probability for the $K = 2$ component of the 
wave function.  This probability is 0.64 compared with 0.002 and 0.35
for the $2^+_1$ and $2^+_2$ levels, respectively.  Thus, the $2^+_3$ state has a
qualitative character as a $\gamma$-vibration but this is diluted by
other components.  This is to be expected for a transitional nucleus such
as $^{152}$Sm.

Important indicators for structure properties are the strengths for intra- and inter-band E2 reduced transition
probabilities. $B(E2;I_i\rightarrow I_f)$'s measured by the Georgia Tech. and Yale collaborations
are shown in Table 1 together with  CHFB+5DCH calculations. As the two sets of experimental data display differences, comparison between B(E2) predictions and measurements necessarily has a global character. The figure of merit of our theory for $^{152}$Sm is as follows : i) the intraband transition strengths have right order of magnitude, especially for the ground state band, ii) the transition between $\gamma$ and ground state as well as between the $\gamma$ and $\beta$ bands are too collective, and iii) the $\beta$ to ground state band transitions display a mixed character. The theory for the $0^+_2 \rightarrow 2^+_1$ transition strength is about 60\% too high. Nevertheless we conclude that the $0^+_2$ excitation is a $\beta$-vibration in $^{152}$Sm.

Our present conclusion is that the structure of the ground state, $\beta$-, and quasi-$\gamma$ bands 
is globally as the 5DCH theory predicts, but there are probably other components in the wave functions, 
such as 2qp excitations and pairing isomerism \cite{Kulp05} that may have an important large effect
 on the out-of-band transitions. More accurate B(E2) measurements that are underway \cite{ca09} 
 will be a valuable asset for making definite statements on the predictive character of present 5DCH 
 calculations and for disclosing which degrees of freedom might be missing in the structure models including the 5DCH one.

%Another important indicator of the relationship between
%bands is the out-of-band transitions that connect them.  
%Experimentally, the $B(E2)$ values of 6 interband transitions
%have been measured.  The largest in both theory and experiment is 
%the $0^+_2\rightarrow2^+_1$ transition, with the theory about 60\% too
%high.  For the other 5 transitions, the theory has only qualitative
%validity: it agrees with experiment on assigning  the
%3 stronger and the 2 weaker transitions but not on the detailed transition
%strengths.
%Our conclusion is that the structure of these states is largely
%as the theory predicts, but there are probably other components in
%the wave function, such as 2 qp excitations of the ground state and pairing isomerism \cite{Kulp05} that
%may have an important large effect on the out-of-band transitions.

%We have also calculated the properties of the third excited $2^+$
%state because it is a simple excitation in the deformed collective
%model, namely the $\gamma$-vibration of the ground state.  Our calculated
%state has the same average $\langle \beta \rangle$ deformation as the ground state,
%supporting a vibrational interpretation.  If it were a $\gamma$-vibration,
%it should have high probability for the $K = 2$ component of the 
%wave function.  This probability is 0.64 compared with 0.002 and 0.35
%for the $2^+_1$ and $2^+_2$ levels, respectively.  Thus, the state has a
%qualitative character as a $\gamma$-vibration but this is diluted by
%other components.  This is to be expected for a transitional nucleus such
%as $^{152}$Sm.

\section{Ground state properties}
\subsection{Nuclear shapes}
\begin{figure*}
\includegraphics [ height= 8cm, angle=-90]{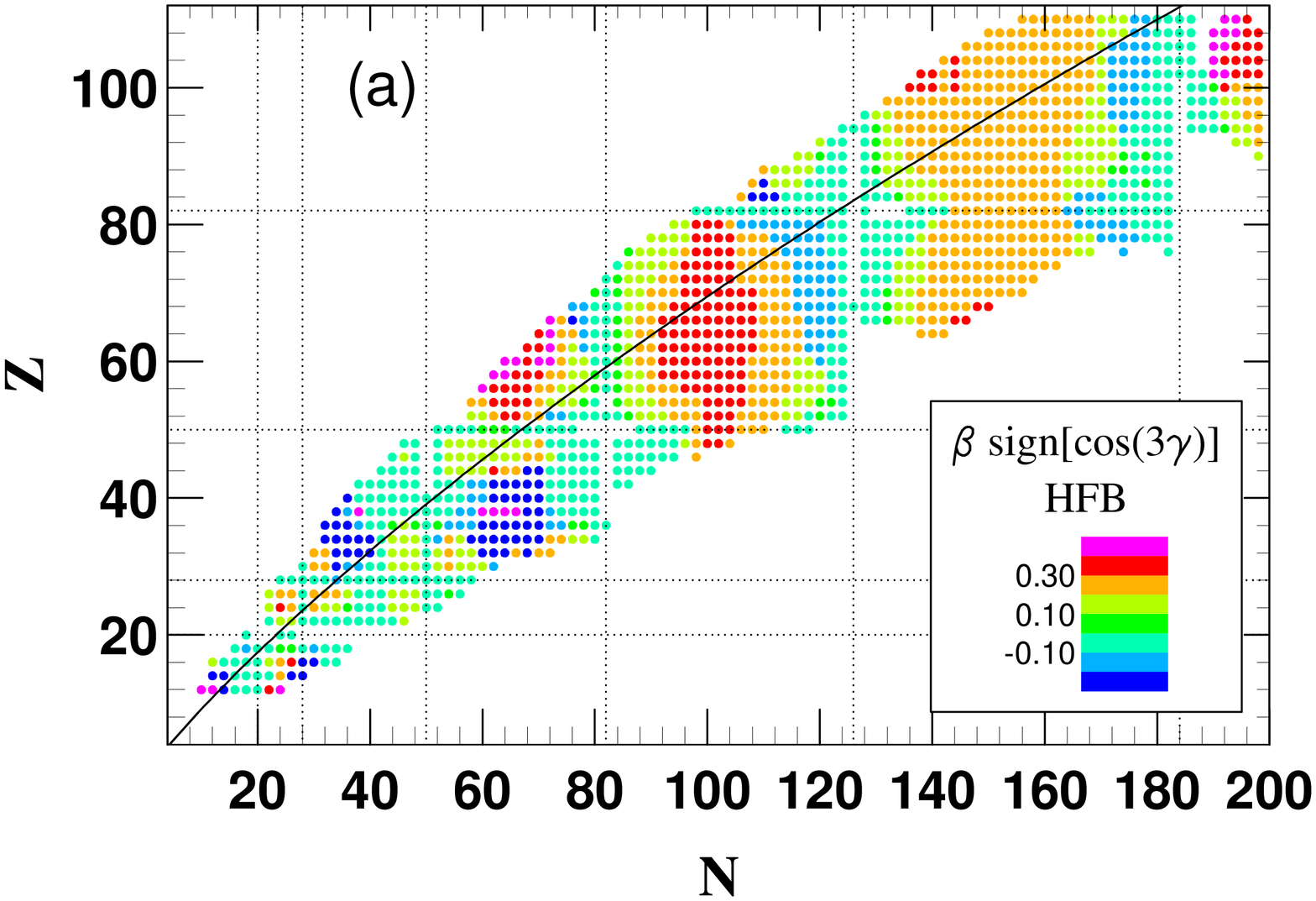}
\includegraphics [ height  = 8cm, angle =-90]{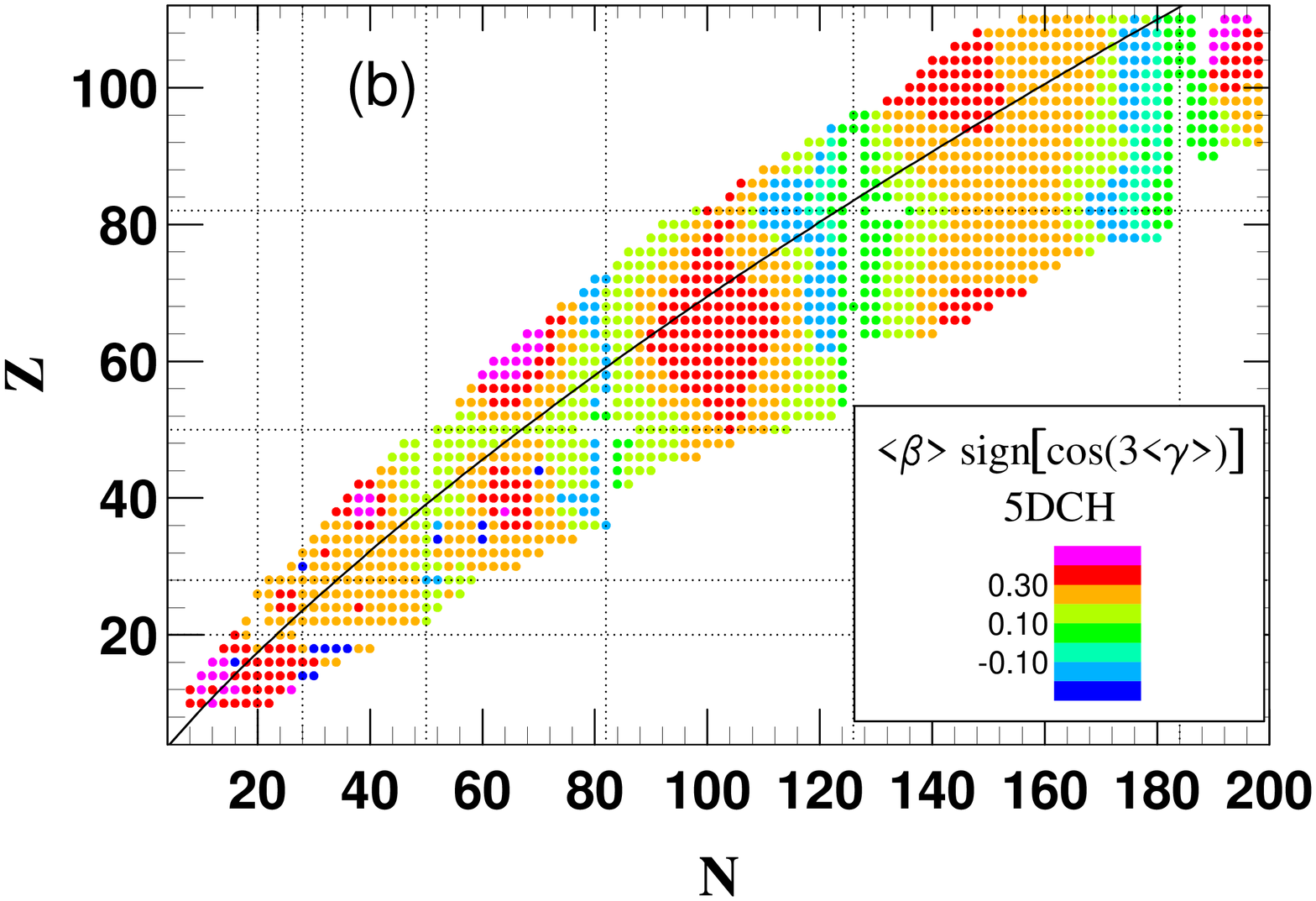}
\caption{\label{fi:chart1}  (Color online) Chart of nuclides showing ground state
deformations.  Panel (a):  HFB minimum;  panel (b): expectation value
in the 5DCH ground state.  The black curve shows the beta-stability line. 
}
\end{figure*}

We begin by displaying in Fig.\ref{fi:chart1} the set of nuclei that we have calculated and included
in our tables.  This comprised
all even-even nuclei that are stable with respect to two-particle emission,
and that have positive correlation energies in the CHFB+5DCH theory. Among the $Z > 96$ nuclei, some ones close to the proton
drip line have been removed from the chart as their inner potential barriers are too low for
inhibiting fission decay. For now, we remark that the
two-particle stability is almost completely determined by the HFB energies.
The correlation energy $E_{corr}$ contribution (see Eq.(\ref{eq:ecorr})) only changes it for a few nuclei 
on the borders.  The next general remark is that the criterion
of positive correlation energy affects only nuclei at magic numbers.  These
are visible as the absence of colored circles along some of the
magic number dotted lines.

In the figure, the color coding shows the deformation $\beta$ of the
ground state, with a $\pm$ sign according to the value of $\gamma$.  
For the HFB ground states, shown in the left-hand panel, 
one sees the familiar landscape of
nuclear shapes, with nuclei near magic numbers having small or vanishing
deformation (dark and medium green), and two large deformed regions 
located at the lanthanides and actinides. Additional regions of 
deformation are centered at nuclei with $(Z,N) = (12,12), (38,40), (40,60)$, and  
$(60,80)$, the heavy nuclei with $N\sim 150$, and the 
superheavy nuclei with $N> 190$. It is also apparent that single magic
numbers do not enforce sphericity. For example, the Sn isotopes are
spherical in the region below $N\sim 82$, become deformed for
neutron numbers in the range $N\sim 100-112$, and get back to spherical 
shapes beyond $N\sim 114$ up to the neutron drip-line.
% Thus the $Z=50$ shell gap collapses at high neutron excess. 

\begin{figure*}
\includegraphics [height = 6cm,viewport=00 01 345
260,clip]{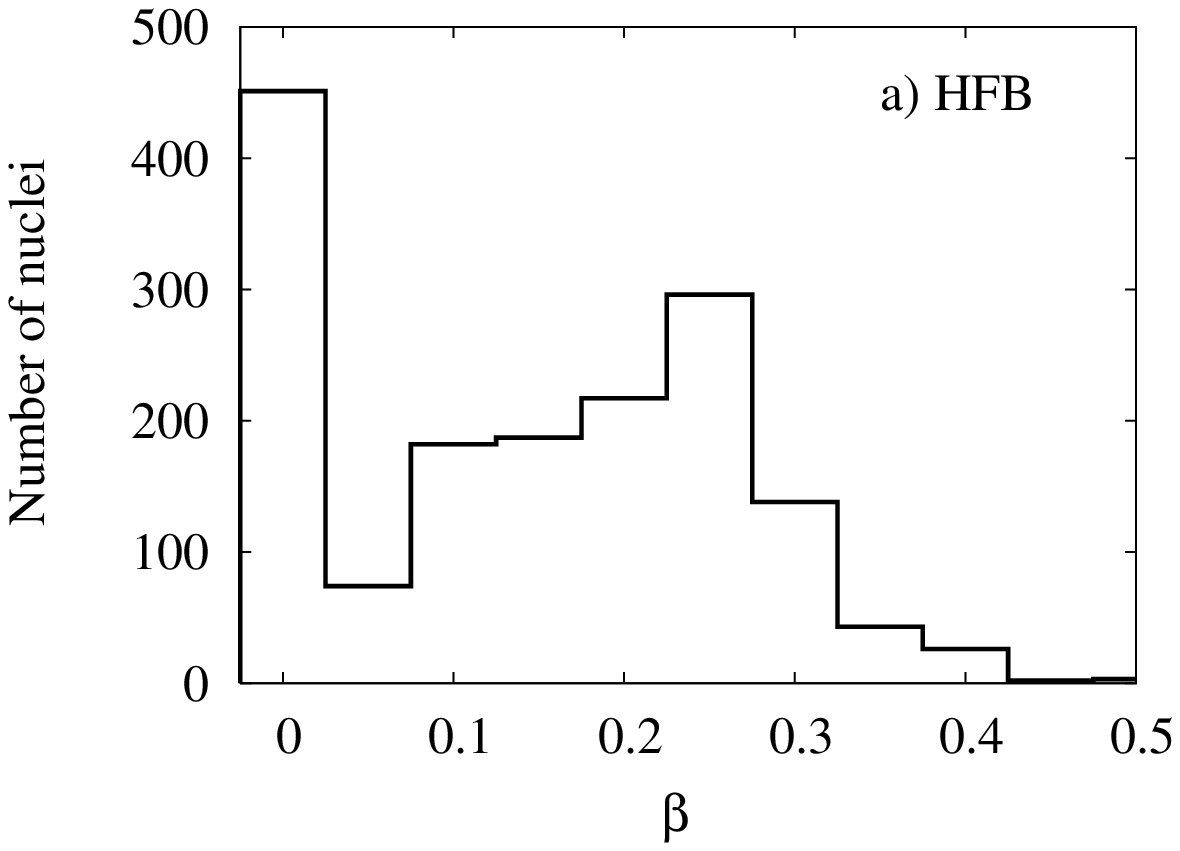}%
\includegraphics [height = 6cm,viewport=69 01 400
260,clip]{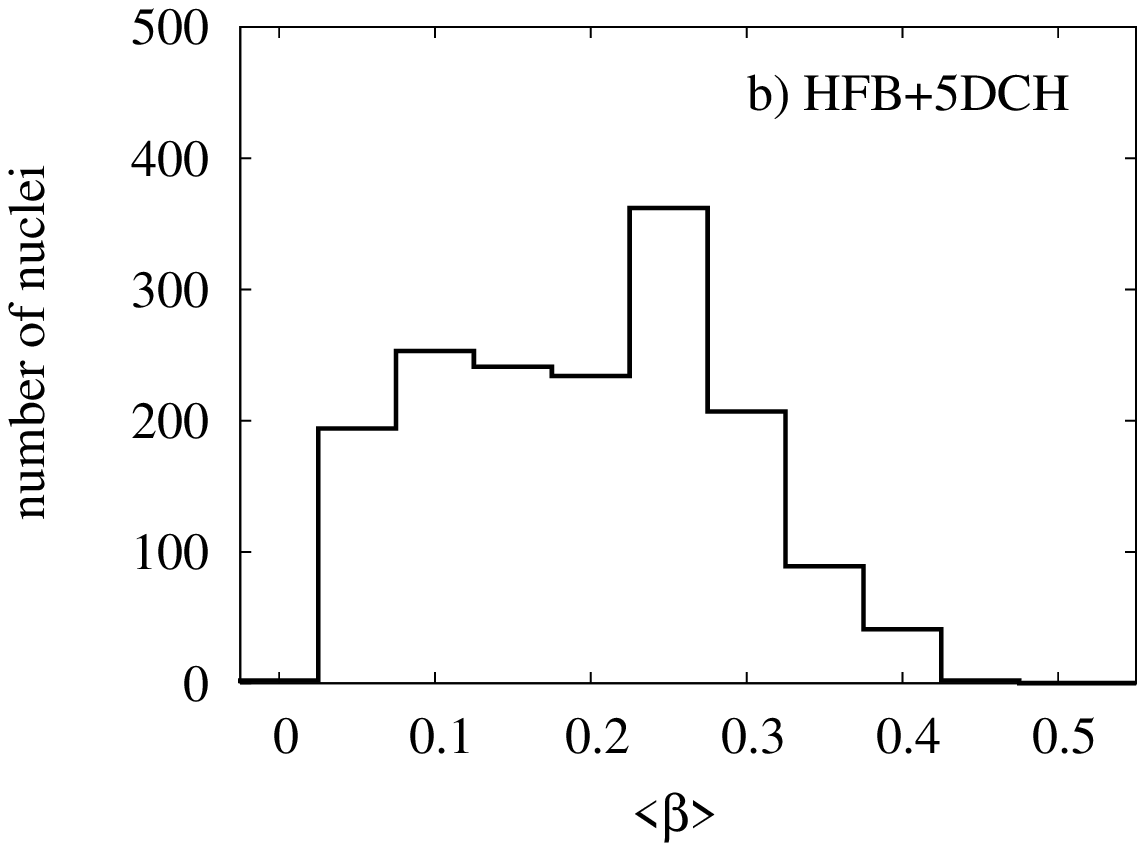}
\includegraphics [height = 6cm,viewport=00 01 345
260,clip]{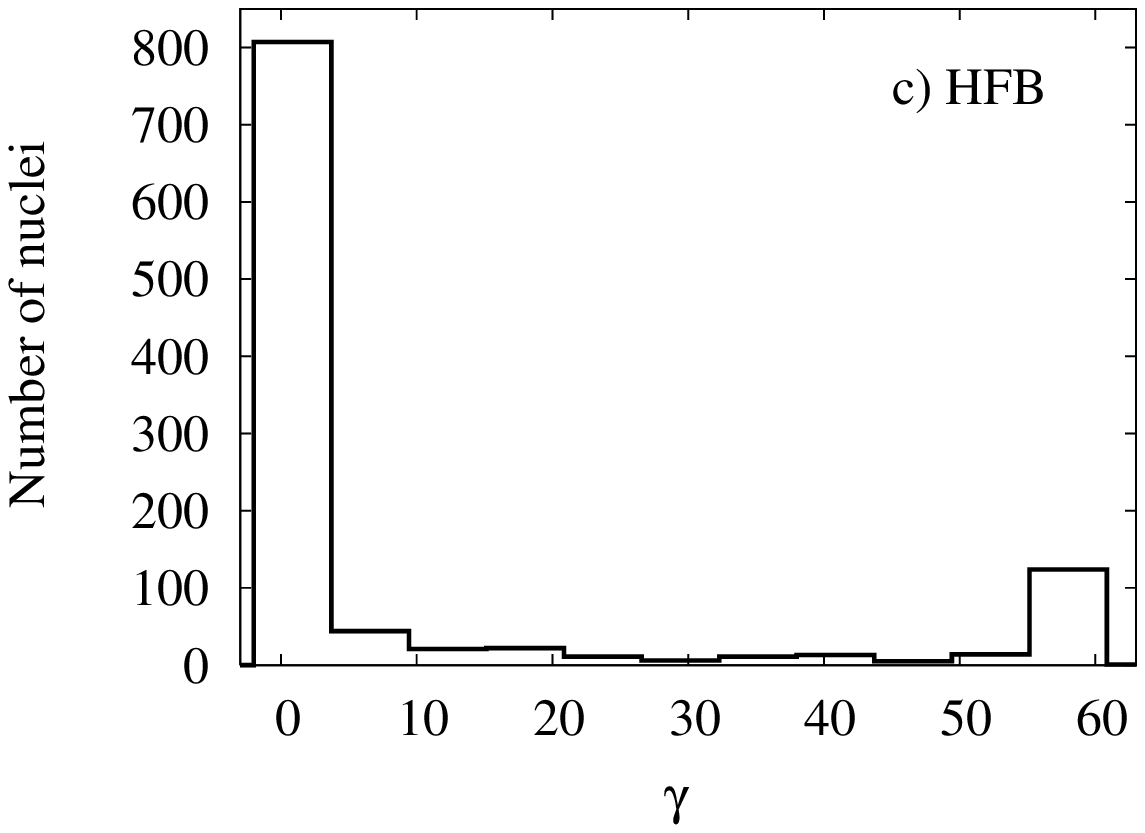}%
\includegraphics [height = 6cm,viewport=69 01 400
260,clip]{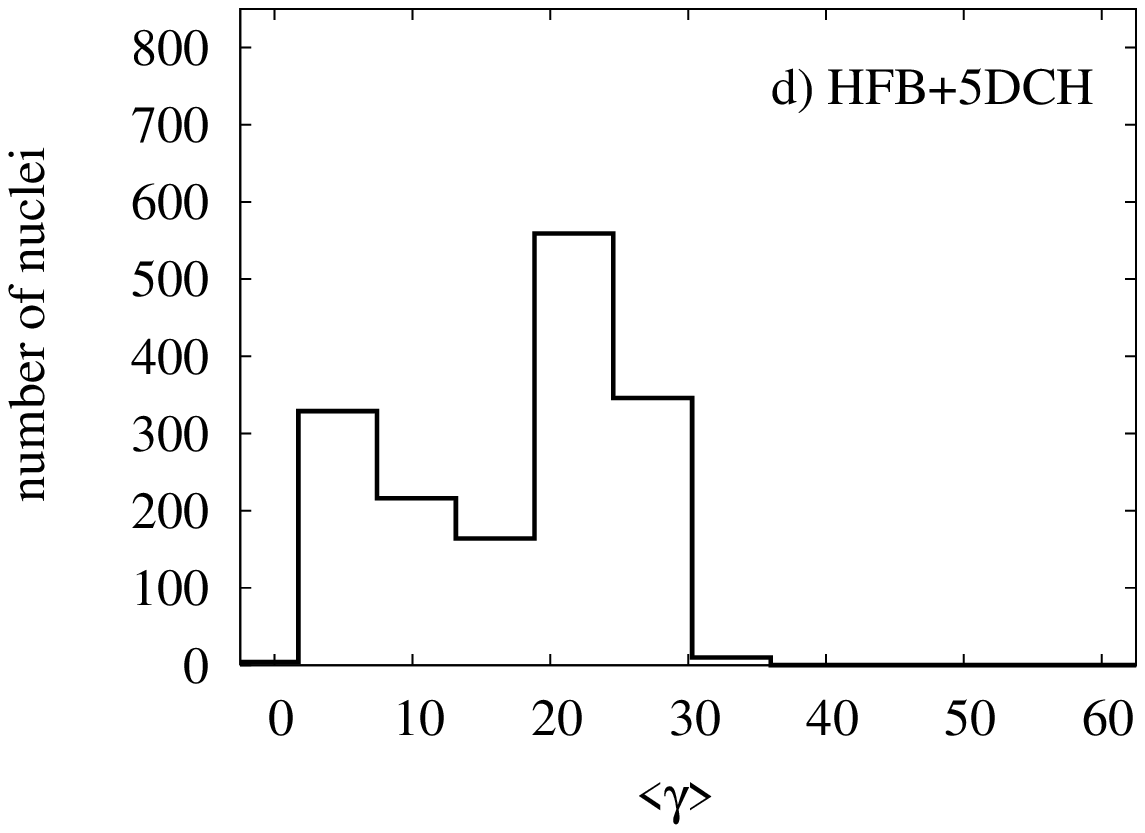}
\caption{\label{fi:bhist}  Distributions of $\beta$ and $\gamma$ in
ground states.  Panels a) and c):
distribution of the HFB minima. Several nuclei at $Z\sim 108$ have
minima at $\beta\sim0.75$ and are not shown.  Panels b) and d): distributions
of $\lb \beta \rb$ and $\lb \gamma \rb$ in the CHFB+5DCH ground state.
The distribution of $\gamma$ in the lower left-hand figure includes only
nuclei with nonspherical minima.  The histogram on lower right-hand
includes nuclei having spherical HFB minima as well.  Units for
$\gamma$ and $\lb \gamma \rb$ are degrees.
}
%int:goutte6/ prune.py  hfb-5dch.dat -> hfb-5dchp.dat
%int:goutte6/ use hist2.py hfb-5dchp.dat 3 0.0 0.7 14   for bmin.hist.eps
% hist2.py hfb-5dchp.dat 16 0.0 0.7 14                  bave.hist.eps
% gamma.py to filter out bmin < 0.1  -->gmin.dat
% hist2.py gmin.dat 4 0.0 80. 14  ->  gmin.hist.dat
% hist2.py hfbp-5dchp.dat 17 0.0 80. 14 --> gave.hist.dat
\end{figure*}

The right-hand panel shows the expectation 
value $\langle \beta \rangle$ for the CHFB+5DCH calculation, with the sign
determined by the expectation value of $cos (3 \langle \gamma \rangle)$.  
The 5DCH wave functions
have larger deformations on average with fewer nuclei near sphericity.
To better see how the 5DCH changes the deformation properties, 
we show in Fig. \ref{fi:bhist} histograms of the distributions of $\beta$ 
and $\gamma$.  The result for the
distribution of $\beta$ in the HFB theory is shown on the upper left-hand
panel.  Among the \Ntot~nuclei in the calculated data set, roughly 30\%
are spherical.  The rest has a broad distribution of deformations peaking
at $\beta\approx 0.25$.  Except for the very heaviest nuclei, 
the largest deformation of the survey was found
for the nucleus $^{24}$Mg, with $\beta=0.54$.  The lower left-hand
panel shows the corresponding distribution of $\gamma$. For this plot,
we restricted the nuclei to those with $\beta> 0.1$, because $\gamma$
is ill-defined in spherical nuclei.  One sees that the great majority
of the nuclei are prolate and axially symmetric, i.e. $\gamma \simeq 0$.  There is 
also a small 
peak for oblate shapes, $\gamma=60^\circ$, comprising about 15\% of the
deformed nuclei.  The paucity of oblate deformations compared to prolate
is well-known in mean-field calculations \cite{ta96,st09}.
However, it should be mentioned again that our calculated nuclei include
only those having positive correlation
energies.  The others are
all near magic numbers and are likely to be spherical.
Turning to the 5DCH $\lb\beta\rb$ distributions shown in the upper 
right-hand panel of
Fig. \ref{fi:bhist}, we see 
essentially all the nuclei become deformed, with deformation broadly 
distributed in the range $0.05\le \lb\beta\rb \le 0.4$.  The corresponding
distribution of axial asymmetries $\lb\gamma\rb$ in the lower left-hand panel shows
that axial symmetry disappears in the 5DCH wave functions, with
average asymmetries going up to $30^\circ$.

\begin{figure*}
\includegraphics [height = 6cm,viewport=00 01 345 260,clip]
{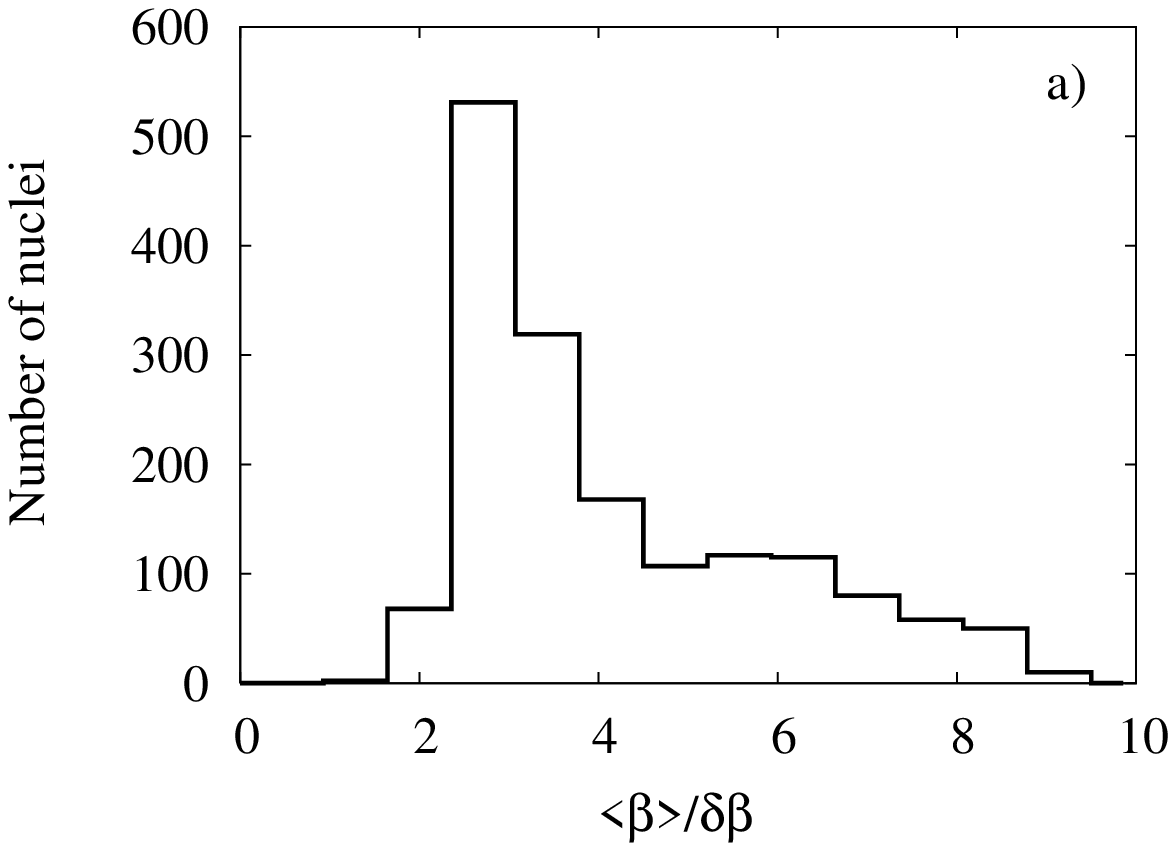}%
\includegraphics [height = 6cm,viewport=00 01 345 260,clip]
{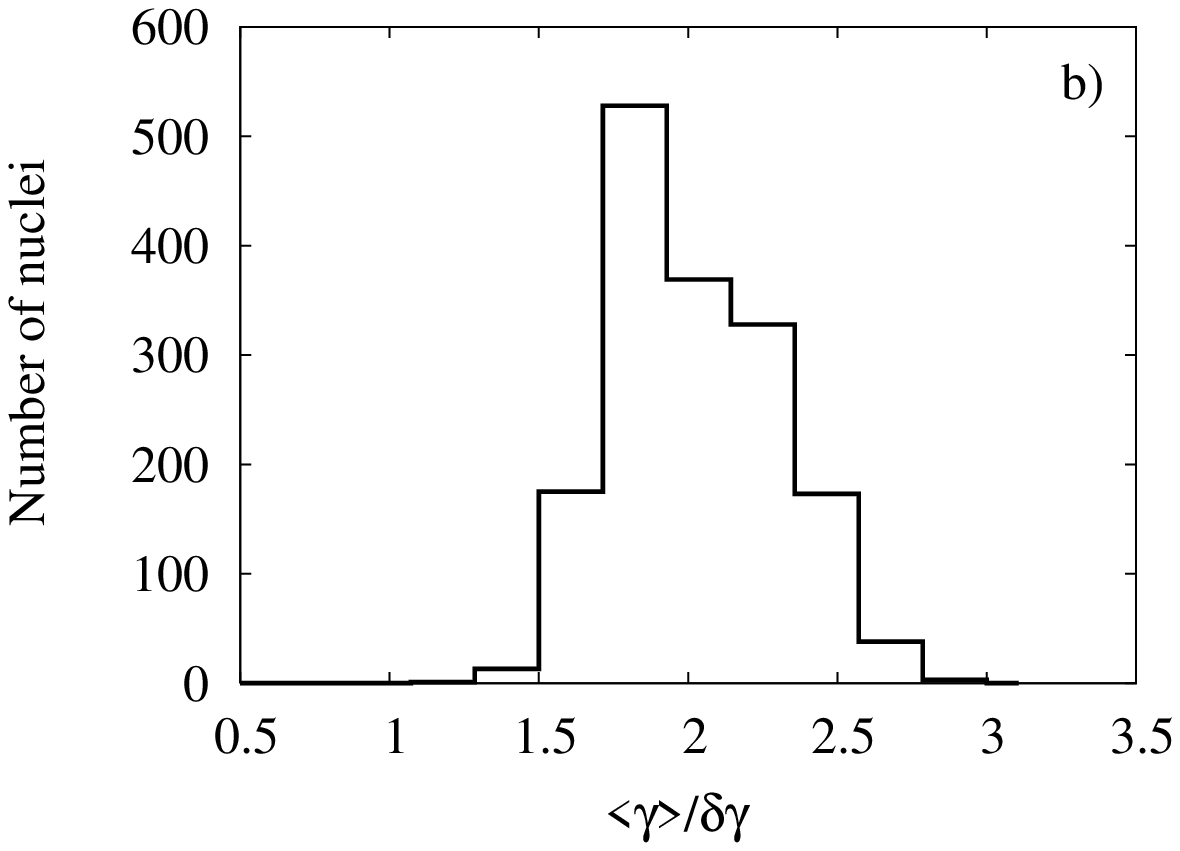}
\caption{\label{fi:rigidity} Distribution of rigidity parameters,
namely
$\lb\beta\rb/\delta\beta$ in panel a) and $\lb\gamma\rb/\delta\gamma$
in panel b).  
}
%goutte6/ rigidity.py *.dat
%hist2.py rigidity.dat 1 0.0 10.0 14
%hist2.py rigidity.dat 2 0.0 3.0 14
\end{figure*}

Additional information about shape fluctuations is provided by
the variances in the deformation parameters, Eq. (\ref{eq:db}). 
In principle, the value
$\gamma=30^\circ$ could arise from a potential energy surface that
is very soft in the $\gamma$ coordinate or from one that has a strong
triaxial minimum.  
Fig. \ref{fi:rigidity} shows the distribution of rigidity measures
$\langle \beta \rangle /\delta \beta $ and $\langle \gamma 
\rangle /\delta \gamma $
for $\beta$ and $\gamma$, respectively.  The $\beta$-rigidity goes to very 
high values, $\langle \beta \rangle /\delta \beta\sim 10$ in the deformed actinides.
We will find that such high values
are present when the nucleus has a well-developed rotational spectrum.
On the other hand, the $\gamma$-rigidity
is much smaller and is never more than $\sim 3$.  Without a clear peaking
at very large values, it will be problematic to
characterize the nuclei in terms of the simple models for triaxial shapes.

\subsection{Radii}
\begin{figure*}
\includegraphics [width = 14cm,viewport=00 42 700 240,clip]{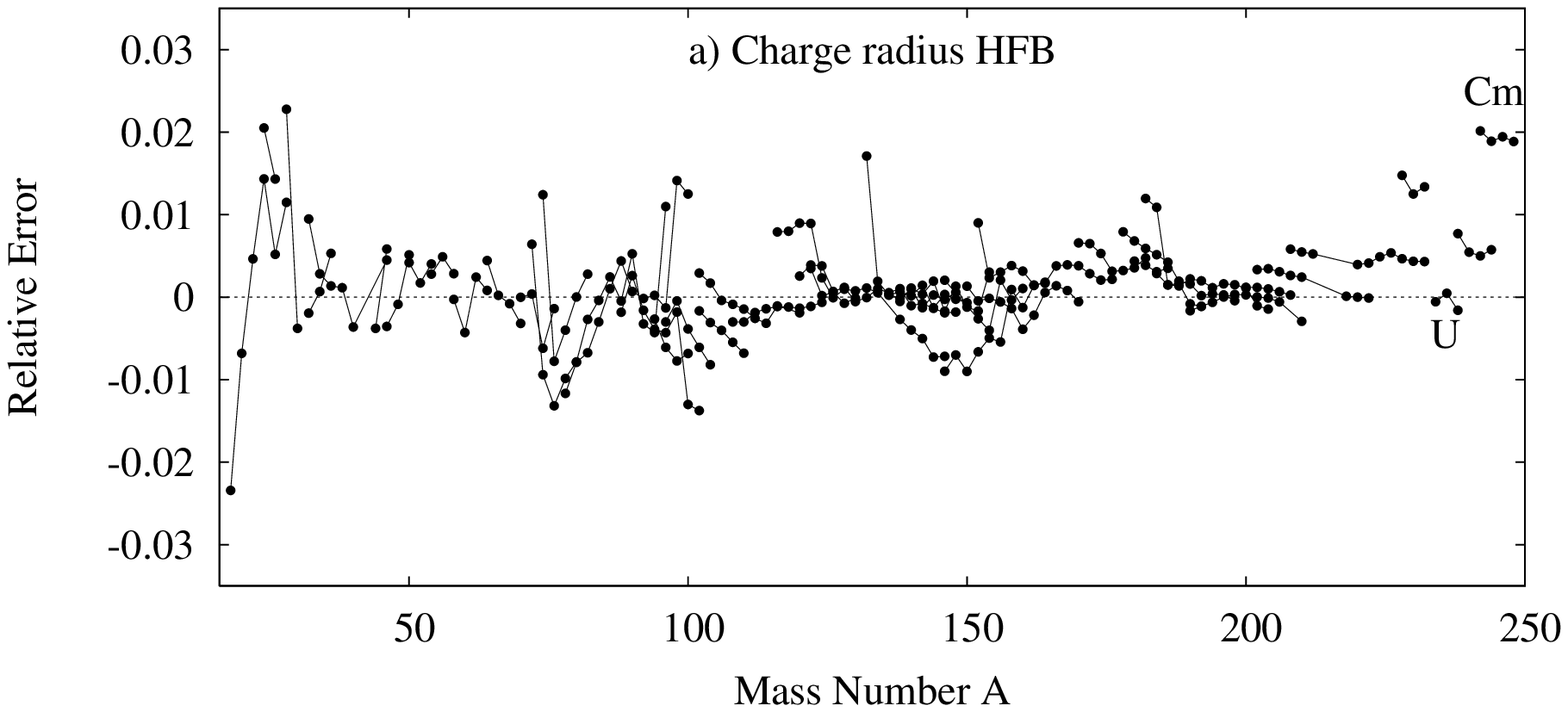}
\includegraphics [width = 14cm,viewport= 00 00 700 240,clip]{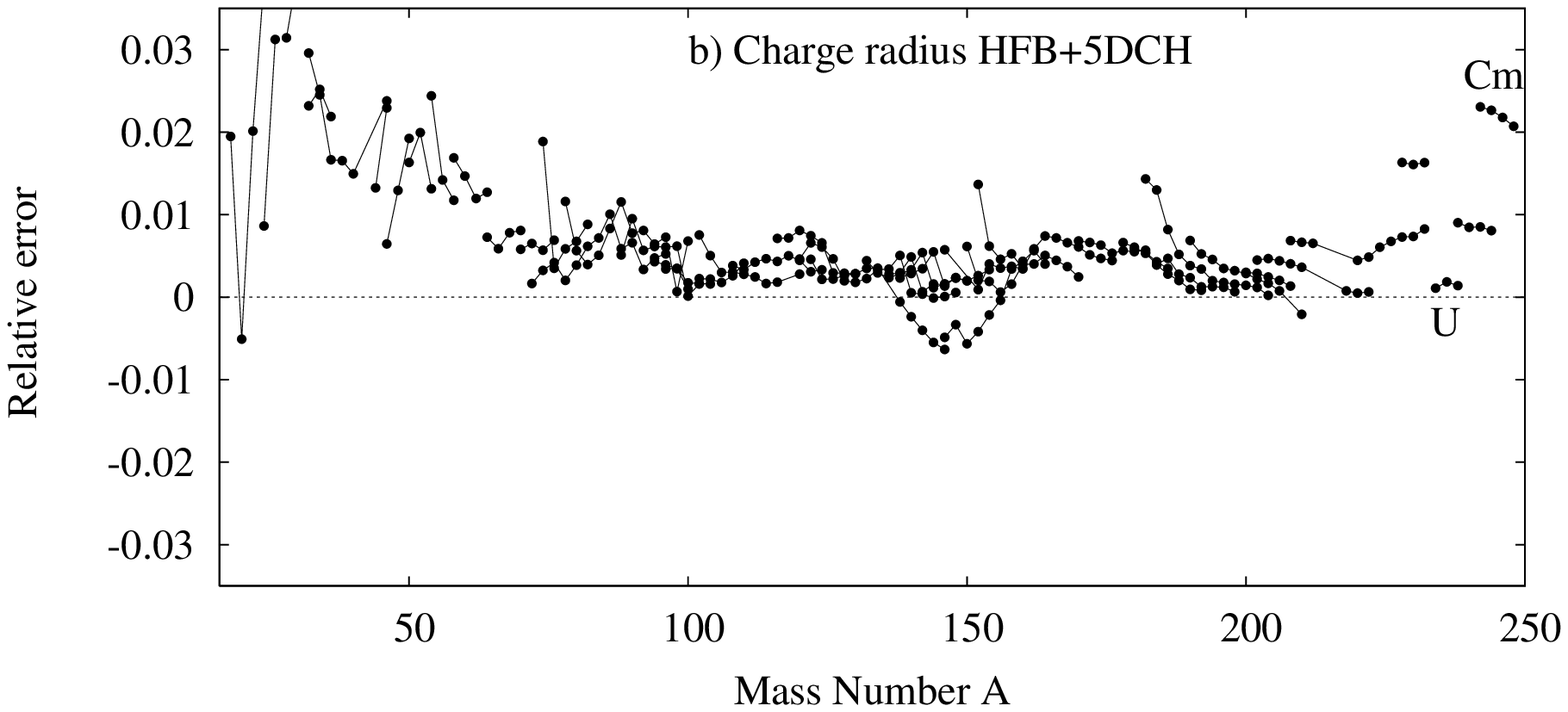}
\caption{\label{fi:radii} Charge radii.  Plotted are the 
relative errors, Eq. (\ref{eq:eps}), with isotopic chains
connected by lines.  Panels a) and b) 
show the results of the HFB and the CHFB+5DCH theories, respectively.
Experimental data is from Refs. \cite{an04,bu00,ca02} (see
\cite{fix}) and includes 313 nuclei.}
\end{figure*}

   We now examine the predicted charge radii, which we compare with
the tabulated experimental data from Refs. \cite{an04,fix}.  
The mean square charge radii $r_{c}^2$ are calculated as \cite{ne70}
\be
r_{c}^2 =  {1\over Z}\int  r^2 d^3 r\, n_p(r) +  r_p^2 +{N\over Z} r_n^2 - 
r_{cm}^2,
\ee
where $n_p(r)$ is the point-proton density, and $r_p^2=0.63$ fm$^2$ 
and $r_n^2=-0.12$ fm$^2$ are the 
rms proton and neutron charge radii, respectively.  The center-of-mass
correction is computed as 
$r_{cm}^2=3\hbar^2/2m \omega A$ fm$^2$ (see Eq.(4.3) in Ref.\cite{ne70}), with $\omega=1.85 + 35.5/A^{1/3}$ MeV. 
%From J.p.'s Email 12.3.2009:
%  <r_c **2> = <r_p **2> +(3/5)(0.65)**2 -41.47/(A omega) -0.12N/Z
% omega =1.85+35.5/(A**1/3)
We show in Fig. \ref{fi:radii} the comparison of calculated and
experimental charge radii, plotted as the relative error 
\be
\label{eq:eps}
\epsilon = r_{c}^{th}/r_{c}^{exp}-1.
\ee  The upper and lower
panels show the HFB and the CHFB+5DCH results, respectively, with 
lines connecting nuclei in isotopic chains.  
We see that the theory is remarkably accurate at the
HFB level, and the CHFB+5DCH hardly changes
the predictions.   Among the heaviest nuclei, we find that the U isotopes
are reproduced very well.  The theory seems to be high for the Cm isotopes,
but it should be noted that these radii were based on systematics in
the absence of any direct measurement \cite{angeli}.

\begin{figure}
\includegraphics [width = 8cm] {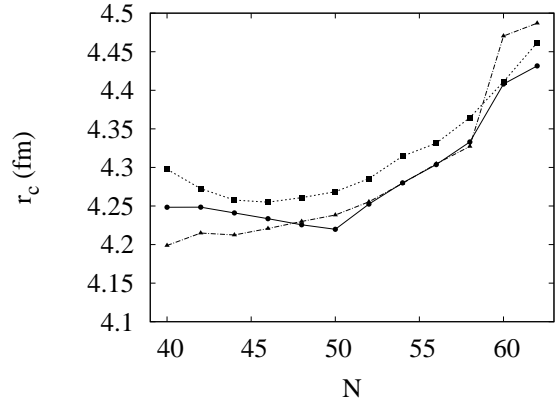}
\caption{\label{fi:zr} Charge radii $r_{c}$ of Sr isotopes.
Experimental: circles joined by solid line; HFB: triangles joined
by dot-dashed line; CHFB+5DCH: squares joined by dotted line.}
%int:goutte6/ r_sr.dat from plot_r.dat; r_sr.gnu
\end{figure}

Nucleus-to-nucleus variations in radii can be attributed to 
deformation changes\cite{otten} as well as other nuclear structure effects
\cite{ta93,re95,sh95}.
The effects of deformation can be easily seen in individual isotopic chains.  An example is the Sr isotopic
chain, shown in Fig. \ref{fi:zr}.  Experimentally, one sees a slight
decrease in
the radius from $N=40$ to the $N=50$ magic number, followed by a
much steeper increase in radii as more neutrons are added.  
The HFB minima are spherical below $N=50$ and deformations increase to very large
values at the heaviest isotopes in the figure. That results
in almost monotone increase in radius
from the lightest to the heaviest isotopes.

Turning to the CHFB+5DCH results, we find that the main effect is in the
lighter nuclei, and it is to increase the charge radius.  This is 
to be expected, since deformations increase the radius and the
average deformations are systematically larger in the CHFB+5DCH.
The largest increase, by 4\%, is in the nucleus $^{30}$Si.
Here
the HFB minimum is spherical, while the CHFB+5DCH ground state has a 
mean deformation $\langle  \beta  \rangle$ = 0.48.
Returning to the Sr isotopic chain, the correlations associated
with the CHFB+5DCH bring the theory in very good overall agreement with data.
This comes about from two effects.  In the very light isotopes,
the CHFB+5DCH predicts large deformations instead of the spherical
shape of the HFB minimum, increasing the radii.  On the other end
of the isotopic chain the nuclei are also deformed, but the average 
deformation in the CHFB+5DCH wave functions ($\langle \beta \rangle \sim
0.3-0.35$) is less than in the HFB minima ($\beta \sim 0.45)$.
 
Table \ref{ta:radii} shows the performance of the theory, using as
a quantitative measure the rms dispersion $\sigma$ about the mean
$\bar \epsilon$,
$\sigma=\langle (\epsilon - \bar \epsilon)^2\rangle^{1/2}$.  Both
HFB and the CHFB+5DCH treatments, suitably renormalized, are accurate to
0.6\%.  For a comparison, the 2-parameter
``Finite surface" model \cite{an04} taking $ r_{c} = r_0 A^{1/3} + r_1 A^{-1/3}$ fm
is shown in the third row of the Table.  Here the error is about 
twice as large.   

\begin{table}
\caption{\label{ta:radii}Comparison of calculated charge radii 
with experiment: $\bar \epsilon$ is the mean of  $\epsilon$ (see
Eq. (\ref{eq:eps}));
$\sigma$ is its rms dispersion  about the average.  313 nuclear
radii were included in the comparison as in Fig. \ref{fi:radii}.
In the column ``HFB (new)" we use the modern value $r_p=0.875$ fm
for the proton charge radius \cite{PDG}.
}
\begin{tabular}{|c|rr|}
\colrule
Theory &     $\bar \epsilon$  & $\sigma$ \\
\colrule
HFB    &  0.001  & 0.006 \\
HFB (new) & 0.005 &   0.007\\
CHFB+5DCH &  0.006  &  0.007     \\
Finite Surface &  0.0000 & 0.012 \\ 
\colrule
\end{tabular}	
\end{table}

\subsection{Correlation energies}\label{corener}
\begin{figure*}
\includegraphics [height = 5.7cm,viewport=00 00 700
240,clip]{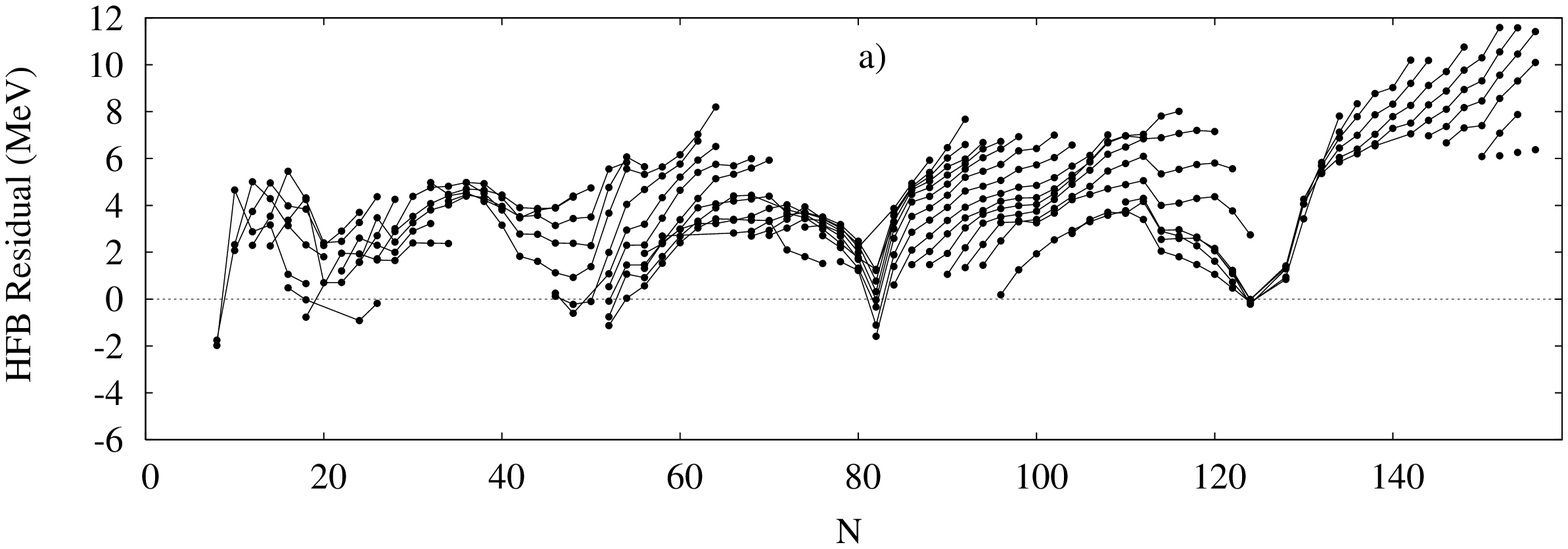}
\includegraphics [height= 5.7cm,viewport= 00 00 700
240,clip]{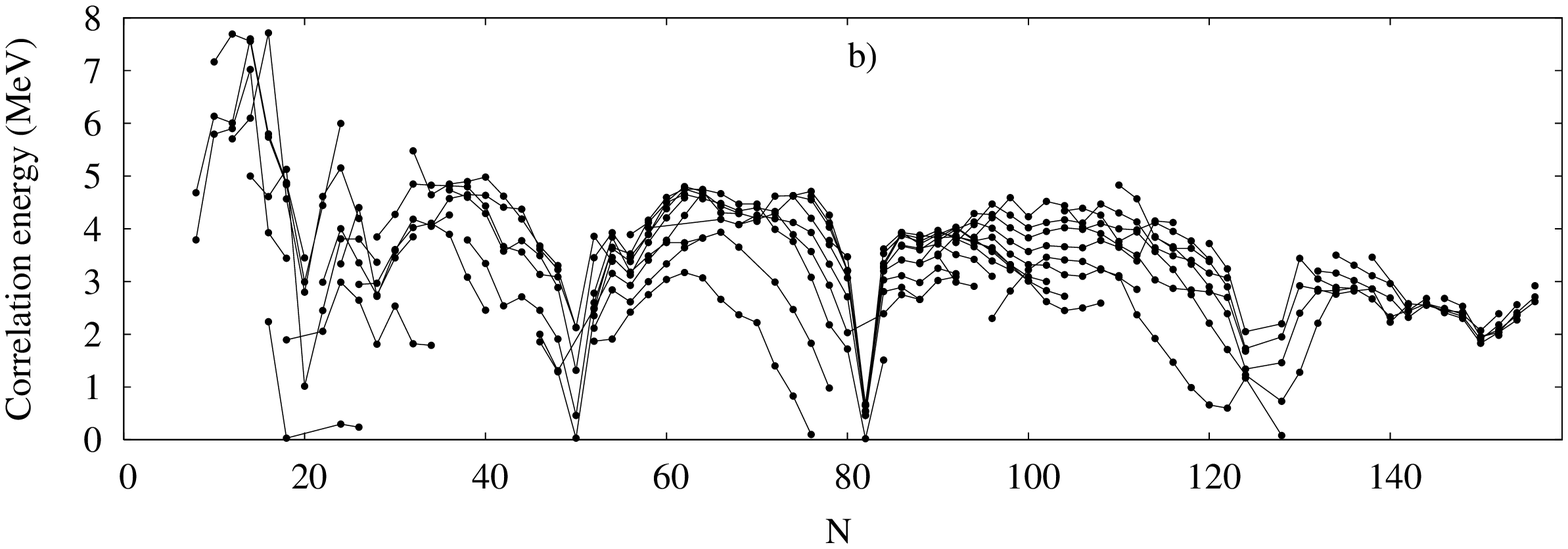}
\includegraphics [height = 5.7cm,viewport=00 00 700
240,clip]{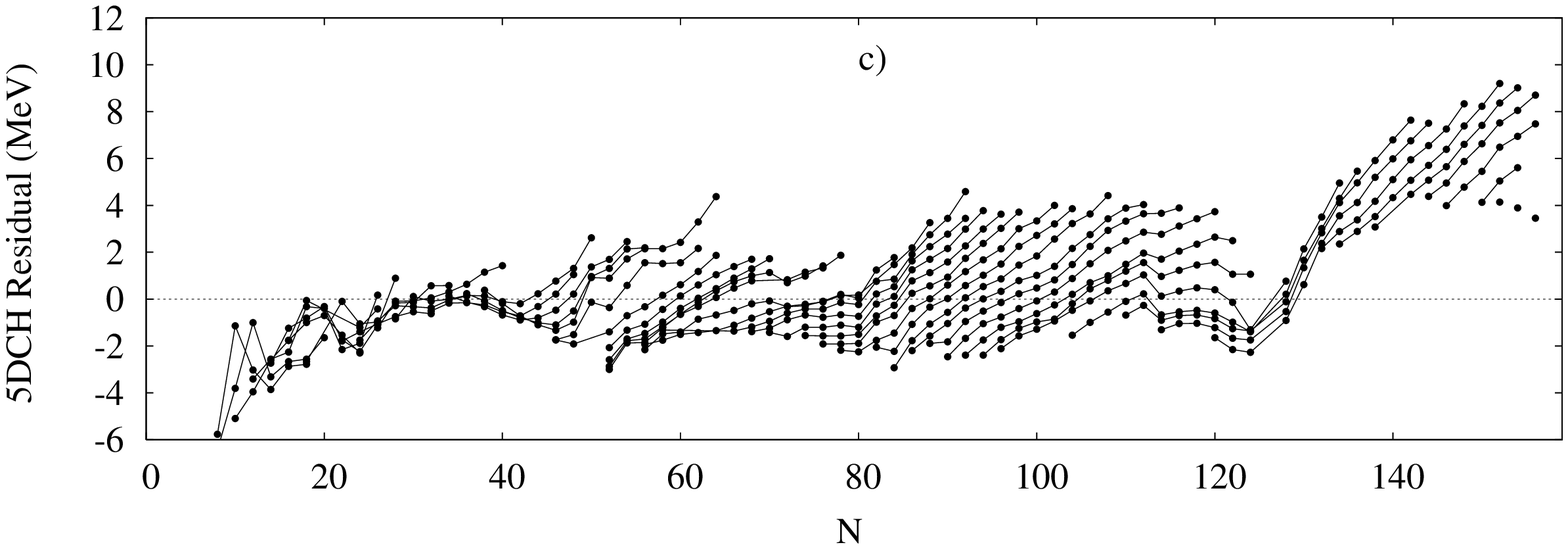}
\caption{\label{fi:ecorr} Panel a):  residuals
of the HFB binding energies with respect to experiment, plotted as
a function of neutron number with nuclei of the same $Z$ connected by
lines.  Panel b):  Correlation energy, Eq. (\ref{eq:ecorr}).  Panel c): residuals including
correlation energies.  Experimental data is from Ref. \cite{Au03}.}
%int:goutte6/  energy.py  hfb-5dchp.dat -> e.dat
\end{figure*}

A key observable that theory should describe is nuclear
masses or equivalently their binding energies. 
We shall consider the binding energy to be composed of two terms,
the binding energy of the mean-field minimum calculated in an unconstrained
(with respect to shape) HFB calculation, and the correlation energy
associated with the spread of the wave function over the quadrupole
shape degrees of freedom. For an orientation, we show in Fig. \ref{fi:ecorr}
these two contributions and their sum, displayed as difference between
experimental \cite{Au03} and theoretical energies (i.e. residuals).  One can see that the shell
effects at $N=82$ and 126 are too large in the HFB theory, and the
correlation energies vary in a way to reduce the shell effects to
a level closer to that needed.

\begin{figure*}
\includegraphics [width = 8cm ] {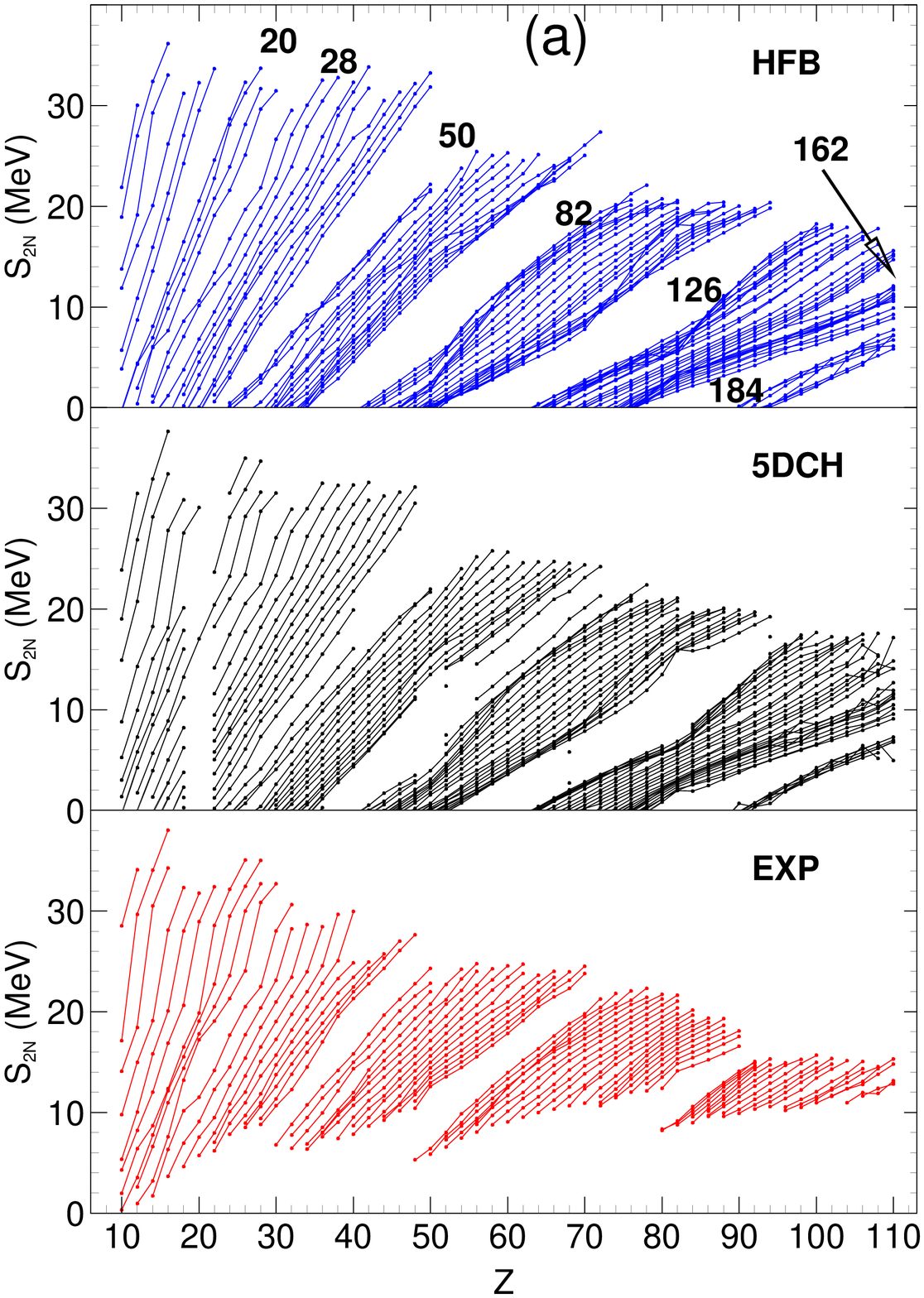}
\includegraphics [width = 8cm] {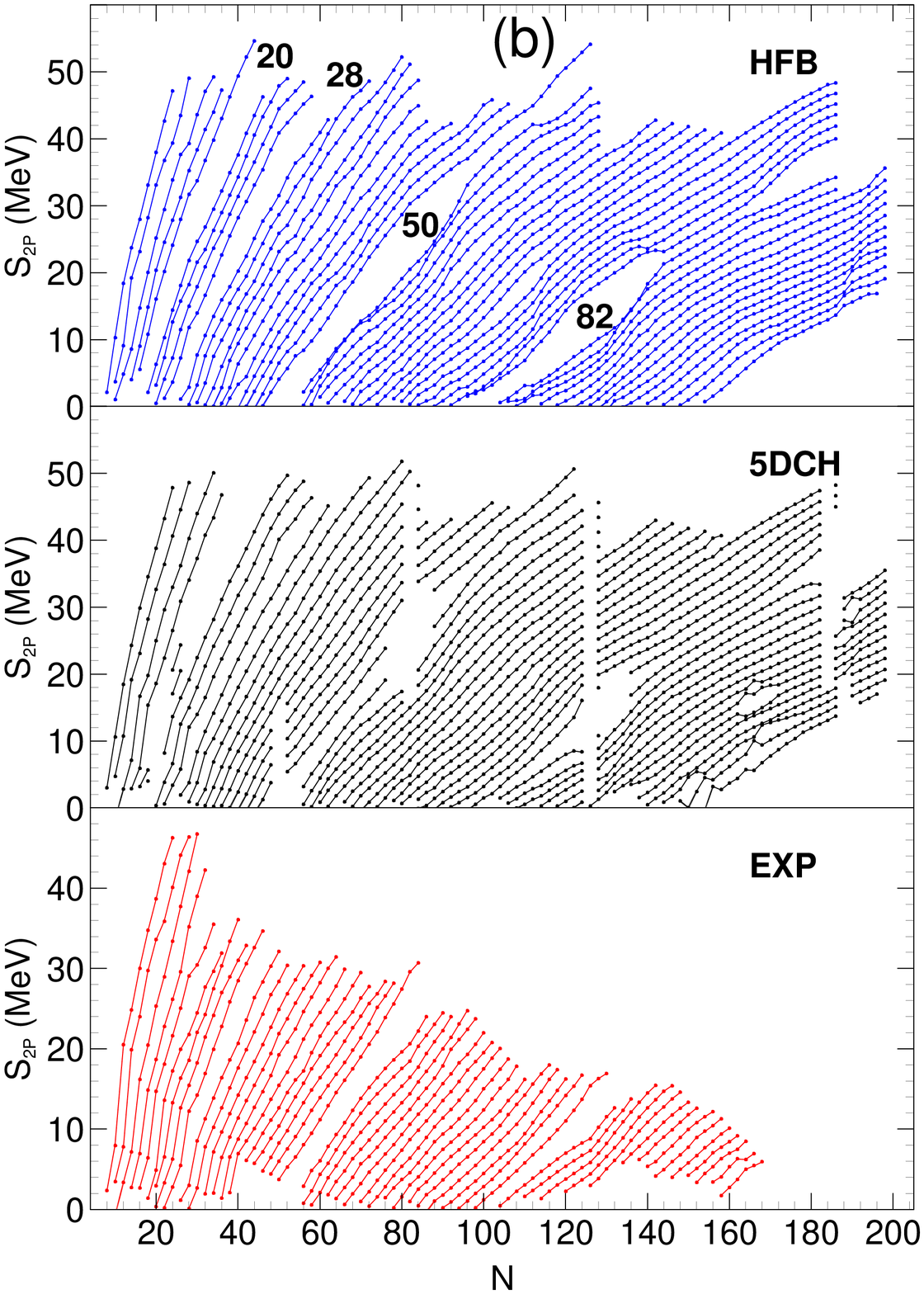}
\caption{\label{fi:s2np}  (Color online)  Panel (a) : two-neutron separation energies;
panel (b): two-proton separation energies. 
The three rows
show the HFB theory, the CHFB+5DCH theory, and experiment, respectively.
Experimental data is from Ref. \cite{Au03}. 
}
\end{figure*}

The overall performance of the theory with respect to masses depends extremely
sensitively on the parameters of the functional, and any useful theoretical
mass table requires that the force parameters be refitted. This has been recently
carried out for the Gogny D1N and D1M parametrizations \cite{hilaire}.  However, in 
the present study we will keep the original D1S interaction
and evaluate the performance with respect to differential quantities,
which are much less sensitive to the precise parameters of the interaction.
The first quantity we examine is the two-nucleon separation energy
defined as
\be
\label{eq:s2n}
\begin{array}{lr}
S_{2n}(N,Z) = E(N-2,Z) - E(N,Z), & {\rm 2\-n~separation~energy},\\
S_{2p}(N,Z) = E(N,Z-2) - E(N,Z), & {\rm 2\-p~separation~energy.}\\
\end{array}
\ee
In the left- and right-hand panels of Fig.\ref{fi:s2np} we show the two-nucleon separation energies 
$S_{2n}$ and $S_{2p}$
for the HFB and the CHFB+5DCH calculations, presenting the calculations in
a similar way as was done in Ref. \cite{be08}.  The shell gaps are
quite obvious, and one can see that they are reduced in the CHFB+5DCH theory.
The available experimental data is shown on the bottom panels.  One
can see that the shell gap varies with the number of nucleons
of opposite isospin.  In particular, it is observed in the right-hand panel for
the proton separation energies that the $Z=50$ and $Z=82$ shell gaps 
disappear at high neutron excess.  As mentioned earlier, the ground states
become deformed in these neutron-rich nuclei. 

In the left-hand panel for $S_{2n}$ one also sees a gradual opening of the
$N=162$
spherical shell gap for proton numbers $Z >96$. This gap 2.5 MeV wide for Z = 110
should increase stability of superheavy elements (SHEs). Our predictions are consistent with those based on calculated
shell correction energies \cite{smol95}, and with the observation of a minimum in alpha-decay energies of SHEs
at $N = 162$ (for a review see \cite{oga07}). Impact of this neutron gap on calculated $S_{2p}$ values is also seen for
$N \simeq 162$ in the right-hand panel. Finally we note that Interacting Boson Model calculations are also supporting evidence 
for a neutron spherical gap in close vicinity of $N = 162$ \cite{bre94}.

Another phenomenon is the
enhancement of the gap near doubly magic nuclei.  This phenomenon, 
called ``mutually enhanced magicity'' \cite{Lu03},  is reproduced much better
by the CHFB+5DCH theory than by the HFB.  The best example is the $Z=82$ gap
of the $S_{2p}$ systematics in the bottom right-hand panel, which becomes
larger near $N=126$.  Unfortunately, the Gaussian Overlap
Approximation does not permit us to calculate the doubly magic nuclei.

In Table \ref{ta:s2n} we show the rms residuals of the calculated 
separation energies
with respect to experiment.  The experimental data is from Ref. \cite{Au03},
including only nuclei whose binding energies are given with experimental
error of less than 200 keV.  As already mentioned, our theory only includes 
nuclei whose correlation energy is positive.  This excludes only about 10\% of the
nuclei in the experimental data set. The number of nuclei in the comparison
is given on the first line of the table.  The first comparison, with
the HFB energies, shows rms residuals of slightly less than 1 MeV for both
separation energies and gaps.  The performance here is slightly better than
was found in the survey based on the Skyrme energy functional Sly4,
reported in Ref.\cite{Ben06}.  The bottom line of the table shows
the energies of the full CHFB+5DCH theory, i.e. with the correlation
energy included.  The improvement is about 25\%.  This is surprisingly comparable
to the results found in Ref. \cite{Ben06}, despite that correlation
energy was calculated in a completely different way.

\begin{table}
\caption{\label{ta:s2n}2-nucleon separation energies and gaps.
Sizes of the compared data sets are given on the first line. Rms
residuals with respect to experiment are given on the third and
fourth lines, for the HFB and CHFB+5DCH theories, respectively.
Energies are in MeV.}
\begin{tabular}{|ll|llll|}
\colrule
\colrule
   & & $S_{2n}$ & $S_{2p}$ & $\delta_{2n}$ &  $\delta_{2p}$  \\
\colrule
Size & theory&455 & 433& 396 & 358  \\
  &  exp. & 492& 467 & 444 & 392 \\
\colrule
Theory &HFB  & 1.00  & 0.91  & 1.06  & 0.98  \\
&CHFB+5DCH & 0.72 & 0.71 & 0.68 & 0.61 \\
\colrule
\end{tabular}
\end{table}

We also carried out the statistics on the two-nucleon gaps.  This
quantity is defined by the next higher order difference,
\be
\label{eq:gaps}
\begin{array}{lr}
\delta_{2n}(N,Z) = S_{2n}(N+2,Z) - S_{2n}(N,Z), &{\rm 2\-n~gap},\\
\delta_{2p}(N,Z) = S_{2p}(N,Z+2) - S_{2p}(N,Z), &{\rm 2\-p~ gap}.\\
\end{array}
\ee
As a particular example, there has been much discussion of evolution
of the $Z=28$ gap for high neutron numbers. 
We find that the CHFB+5DCH energies are
below the HFB values, thus weakening any shell effect at
$Z=28$.  There is a peaking at $N=28$ that could be attributed
to ``mutually enhanced magicity'' or to an $Z=N$ symmetry effect, the ``Wigner energy''. 
Experimentally, there is a slight peaking in the gap at $N=40$, 
but we find that it is smooth in the CHFB+5DCH theory.

The overall statistics for the performance of the theories
with respect to two-nucleon gaps are also shown in Table \ref{ta:s2n}. 
The results are somewhat better than those for the separation energies.

%%%%%%%%%%%%%%%%%%%%%%%%%%%%%%%%%%%%%%%%%%%%% 
\section{Yrast spectrum}
In this section we report the predictions for the lowest excitations
of angular momentum $J=2,4$ and $6$.  
For the quantitative measure of the global performance of the theory,
we will use the same figures of merit as in Ref. \cite{be07}.  Because the quantities 
span a large range values, we examine the statistics of the logarithmic
ratio of theory to experiment, namely
\be
R_x = \log(x_{th}/x_{exp}),
\ee  
for a quantity $x$.  We present its average over the data set $\bar R_x$
as well as the dispersion about the average, 
\be\sigma_x\equiv
\lb (R_x -\bar R_x)^2\rb^{1/2}.
\ee   
The results for the properties we can compare with tabulated 
experimental data are discussed individually below and in Sec.
\ref{nonyrast}, and are summarized in Table
\ref{ta:metrics} in Sec. \ref{summary}.

\subsection{The first $2^+$ excitation}
The first physical property we examine is the fraction of  the
energy-weighted sum rule (EWSR) contained in the $2^+_1$ excitation.  
That quantity
is governed more by the inertial and mass properties of the CHFB+5DCH 
than by the topology of the potential
energy surface.  The sum rule fraction is often expressed 
with respect to Lane's isoscalar sum rule ~\cite{Lane}
\be 
S(I) = \sum_i E(2^+_i) B(E2; 0^+_1  \rightarrow 2^+_i)={25\over 4\pi }(\frac{\hbar^2}{m}) A
\langle r^2 \rangle ,
\ee
where $m$ is the nucleon mass and $\langle r^2 \rangle$ is the mean square
mass radius. It is also common to make the approximation $\langle
r^2\rangle= 1.2^2 A^{2/3}$ fm$^2$ \cite{ra01} but we shall rather use our  
calculated mass radius. 
The charged part of the isoscalar sum rule is derived
from $S(I)$ assuming that the charge current and mass current are proportional~\cite{ra01}
\be 
S(II) = S(I)(\frac{Z}{A})^2.
\ee
The fraction s(X) of the sum rules carried by the $2^+_1$ excitations is calculated as
\be 
s(X)= E(2^+_1) B(E2; 0^+_1 \rightarrow 2^+_1)/e^2 S(X), \,\,\,{\rm with}
\,\,\,X=I,II.
\label{eq:sumrule}
\ee

\begin{figure*}
\includegraphics [width = 8cm]{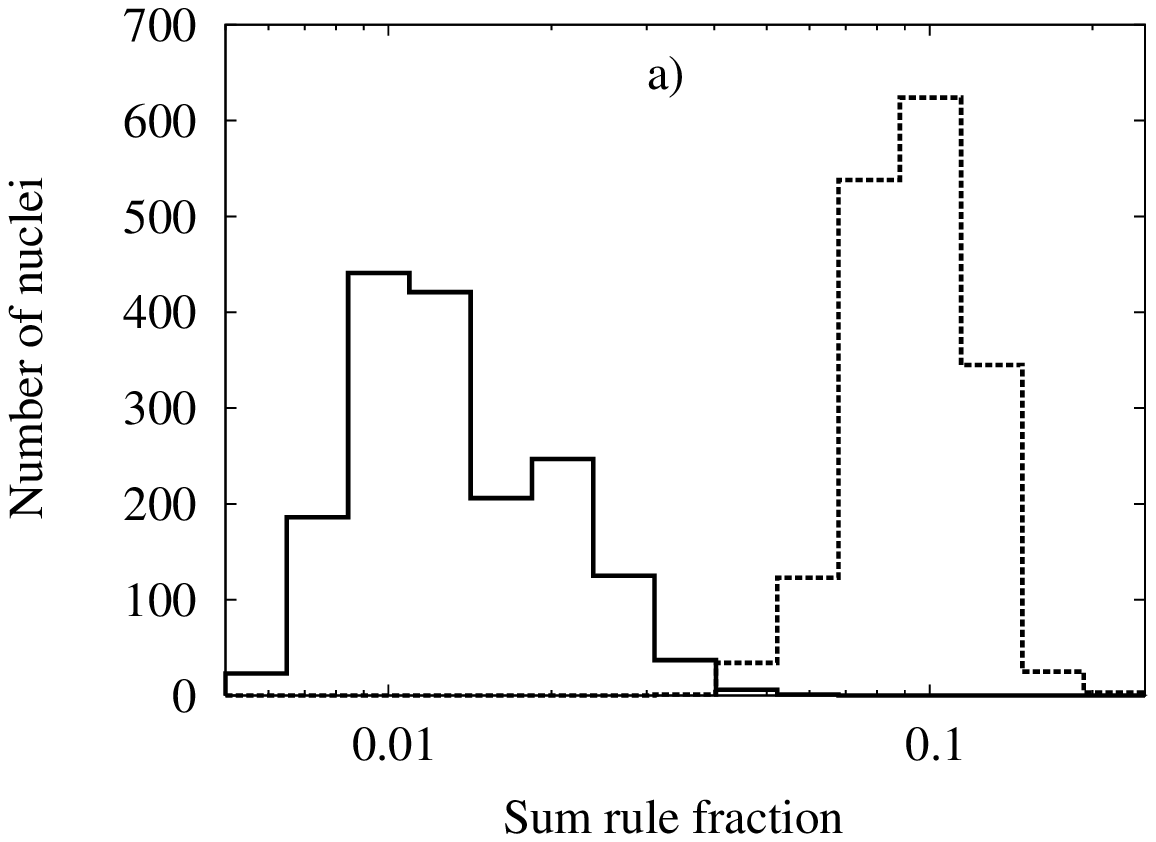}%
\includegraphics [width = 8cm]{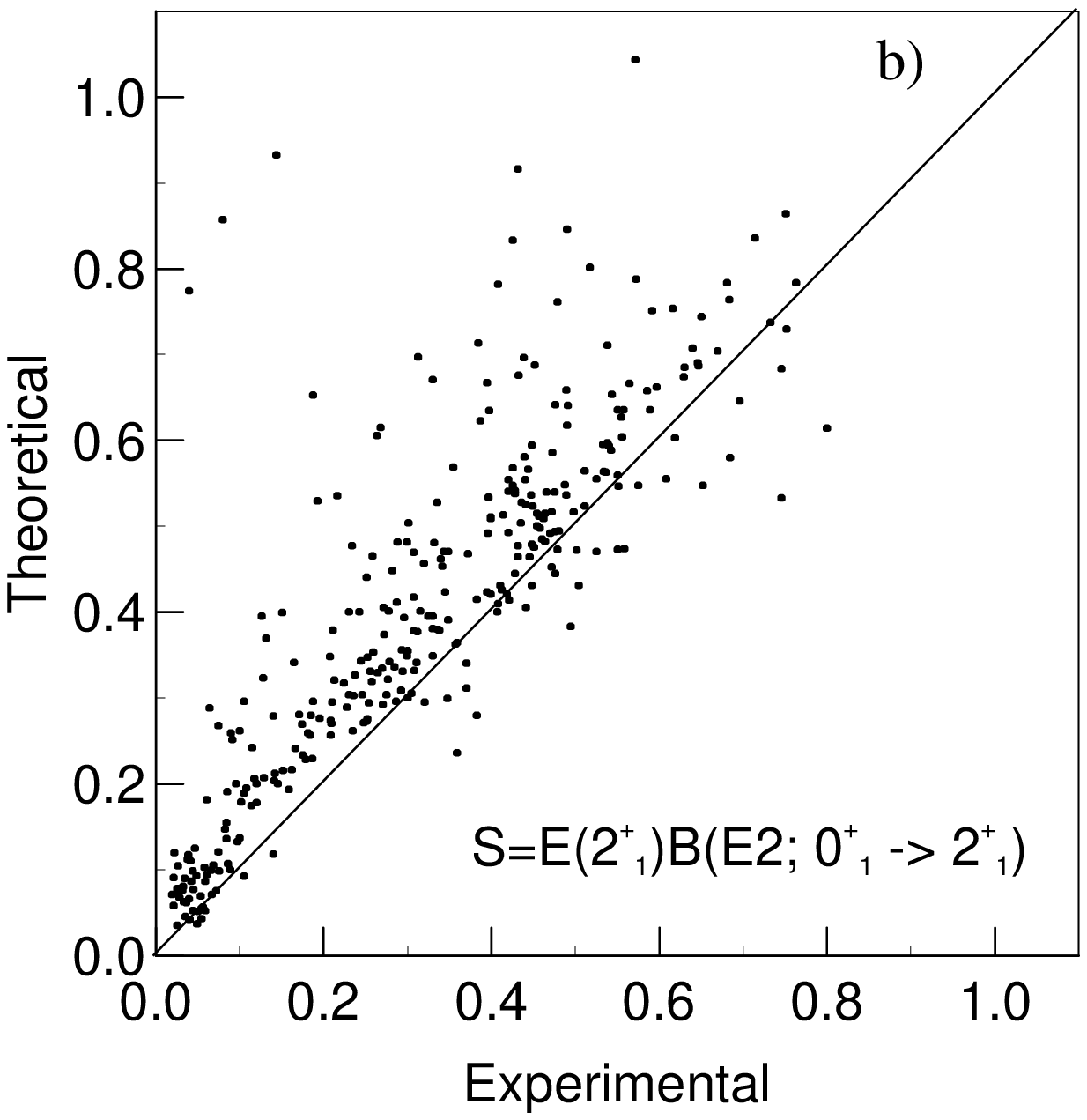}
\caption{Panel a): distribution of sum rule fraction $s(X)$, Eq. (\ref{eq:sumrule}),
in the CHFB+5DCH theory, for the \Ngoods calculated nuclei.  Solid and dashed 
lines show the fraction of the S(I) and S(II) sum rules, respectively.
Panel b): calculated $S \equiv E(2^+_1) B(E2; 0^+_1 \rightarrow 2^+_1)$ versus experimental, for 311 nuclei.  Experimental data
are from Refs. \cite{ra01,sh94}. S values are in MeV-$e^2b^2 $ units.  
}
\label{fi:sumrule}
\end{figure*}

Histograms of s(I) and s(II) are shown on the left-hand panel of 
Fig. \ref{fi:sumrule}.
Individually, the excitation energies and transition strengths vary over several
orders of magnitude.  But their product scaled by $s(X)$ compresses the
rms variation down to about a  factor of 2.  This may be seen in the
histograms of $s(I)$ and $s(II)$ shown in the left-hand panel of 
Fig. \ref{fi:sumrule}.   The fraction of strength in each sum rule
is about  1.5$\%$ for $S(I)$ and 10\% for $S(II)$.
As well known, most of the strength is carried by the 
giant quadrupole resonance.
One can also see from the histograms that the scaling with $S(II)$ produces
more compressed distribution than scaling with $S(I)$.
A scatter plot of the
theory versus the experimental values of $S \equiv E(2^+_1) B(E2; 0^+_1 \rightarrow 2^+_1)$ is shown on the right-hand
panel of Fig. \ref{fi:sumrule}.  There is a concentration of points
on the diagonal that show very good agreement; these mostly correspond
to strongly deformed nuclei.  Overall, the theory somewhat overestimates
the fraction of the EWSR carried by the $2^+_1$ state.

\begin{figure}
\includegraphics [width = 8cm]{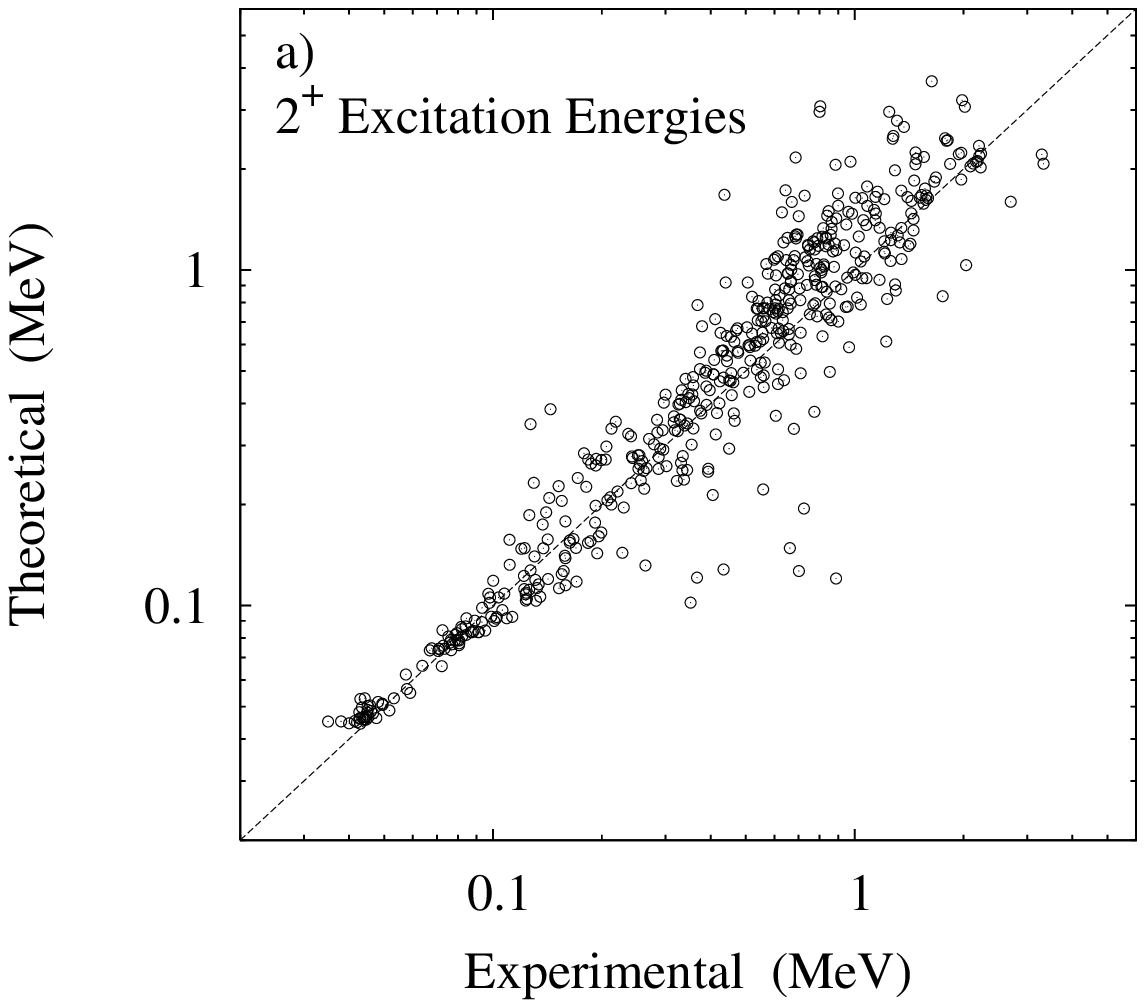}
\includegraphics [width = 8cm]{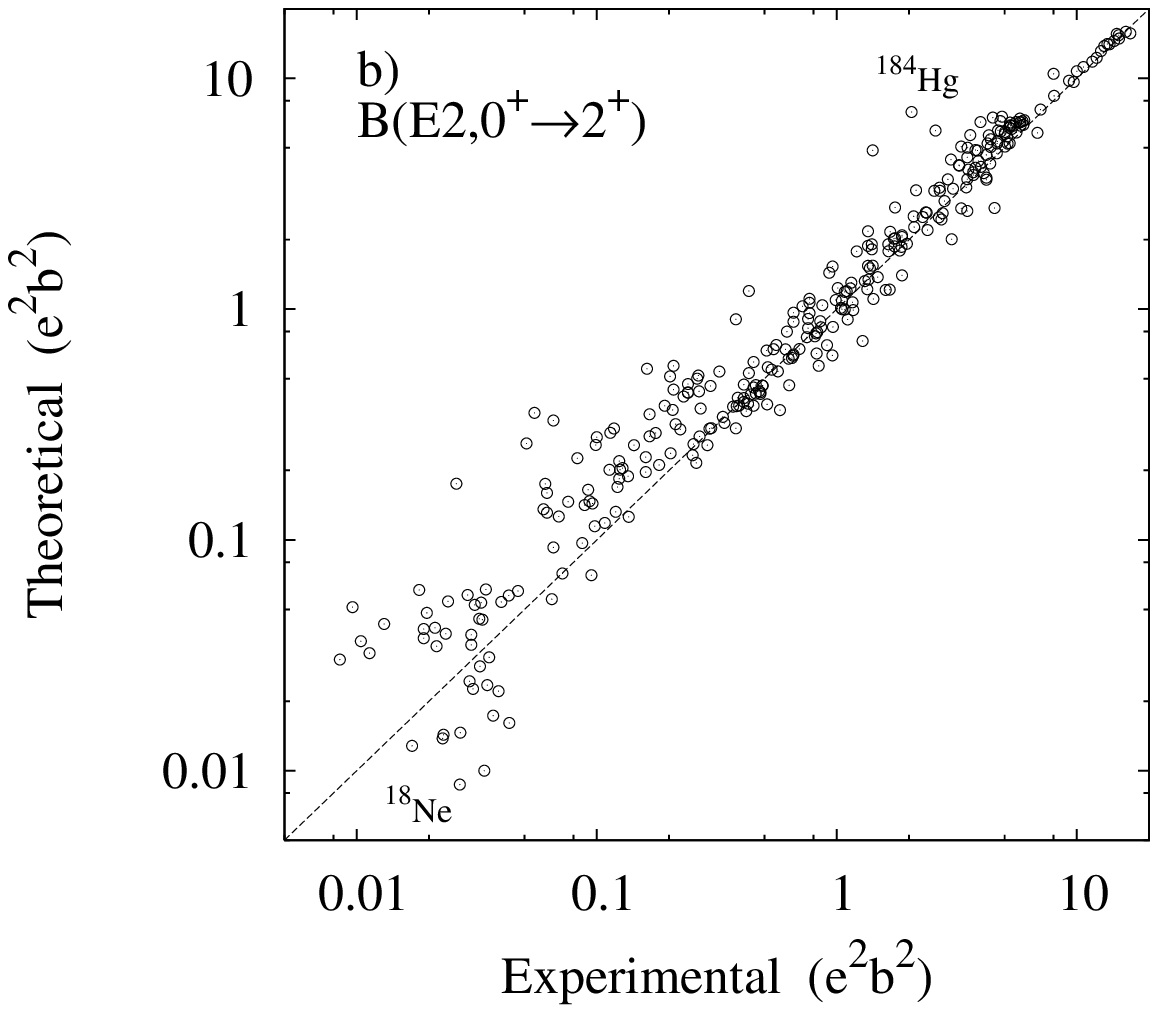}
\caption{Panel a): theoretical $2^+_1$ excitation energies of 537 even-even nuclei 
as a function of their
experimental values. Panel b): theoretical $B(E2; 0^+_1\rightarrow 2^+_1)$
transition strengths of 320 even-even 
nuclei 
as a function of their
experimental values. Several cases showing deviations are labeled by the
nucleus.  Experimental data is from Refs. \cite{ra01,sh94}.}
\label{fi:eq_scatter}
\end{figure}

We now turn to the comparison of excitation energies and transition
strengths with experiment.  The results were reported already in Ref.
\cite{be07}, but we since discovered that the code we had been 
using to solve the 5DCH did not have the desired precision for 
smallest excitation energies.
In the present work we report recalculated energies using a more
accurate code described in Refs. \cite{li82,Li99}.
The comparison of experiment~\cite{ra01,sh94}
and the calculation
is shown in the left-hand panel of Fig. \ref{fi:eq_scatter}.  
The points at the lower left correspond to the deformed lanthanides
and actinides, and one sees that the theory does very well there.
Right-hand panel of Fig. \ref{fi:eq_scatter} shows a similar comparison 
for the
$B(E2; 0^+_1 \rightarrow 2^+_1)$ transition strength.  The points on the upper right side of the figure
correspond to the very deformed actinide nuclei.  Again, the theory
is seen to be remarkably accurate under the conditions of a large
static deformation. 
The global performance figures of merit for the $2^+_1$ energy and the 
$B(E2;0^+_1\rightarrow 2^+_1)$  strength
are given on the first two lines of Table \ref{ta:metrics} posted in Sec. \ref{summary}.

\subsection{The first $4^+$ excitation and  $R_{42}$}
\begin{figure}
% figure is from gene:casten/r42.eps
\includegraphics [width = 8cm]{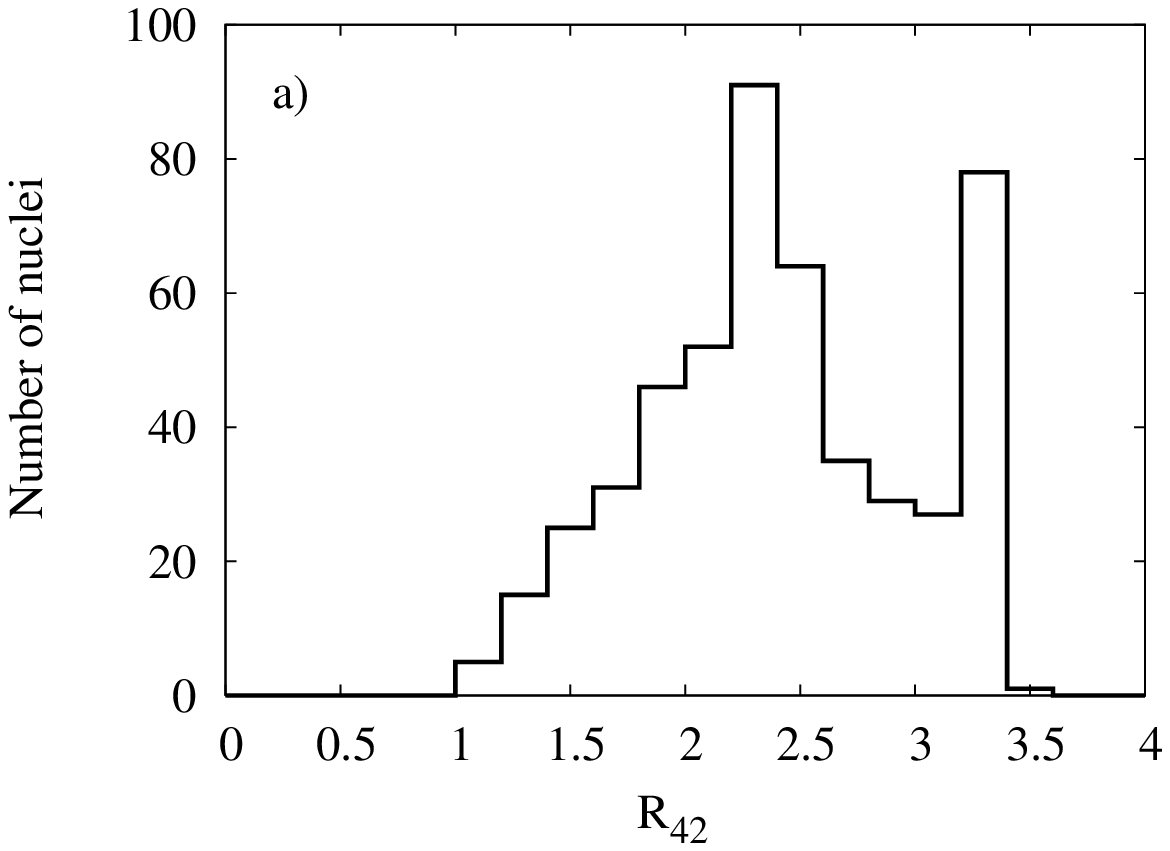}
\includegraphics [width = 8cm]{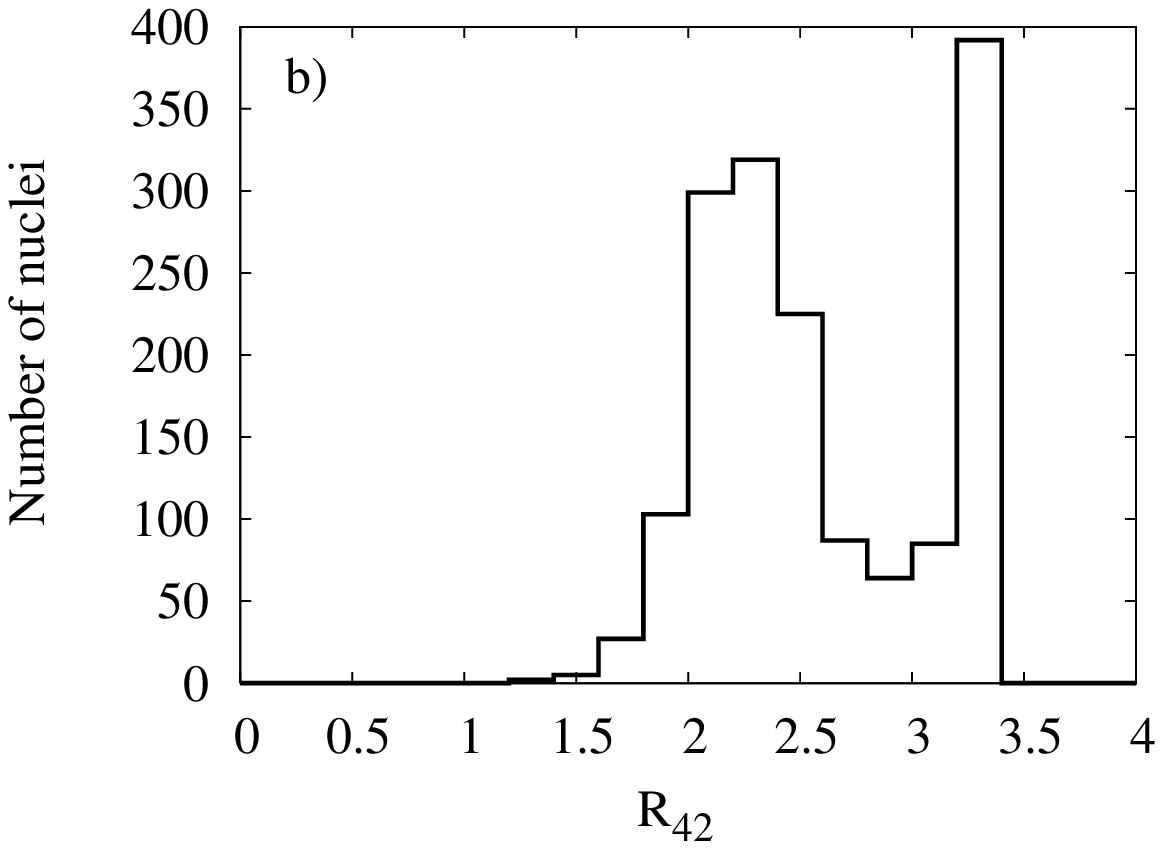}
\caption{Panel a): histogram of experimental $R_{42}$ ratios, 
Eq. (\ref{eq:r42}), for 
501 even-even nuclei, with data from Ref. \cite{brookhaven}. Panel b): 
histogram of calculated $R_{42}$ ratios for \Ngoods even-even 
nuclei calculated in the CHFB+5DCH theory.
}
\label{fi:r42}
\end{figure}

An important signature of the character of the excitation spectrum is
the relationship of the lowest $4^+$ excitation and the $2^+_1$ below it.
A very useful indicator is the ratio of the two excitation energies $E(J^\pi_n)$,
\be
\label{eq:r42}	
R_{42} = {E(4^+_1) \over E(2^+_1)}.
\ee
The $R_{42}$ indicator has been much used, particularly in discussing
complex spectra.   The value $R_{42}= 10/3$ is characteristic of an axial rotor,
$R_{42}=2$ of a vibrator, and $R_{42}=5/2$ of a $\gamma$-unstable
rotor or the $O(6)$ 	
algebraic model \cite{ca00}.  In Fig. \ref{fi:r42} we
display histograms of the experimental and theoretical ratios side
by side.  One sees a very narrow peak at 10/3, 
showing that one can make a nearly unambiguous assignment of axial 
rotors.  In the algebraic
models the three simple limits mentioned above represent extremes
in the parameter space of the models, and it is interesting to see
which ones are favored in the global systematics.
While the axial rotor is clearly special, neither the harmonic vibrator nor
the $\gamma$-unstable rotor shows a corresponding accumulation
in the experimental data.  The CHFB+5DCH theory, on the other hand, does
show a second peak just below the $\gamma$-unstable value, $R_{42}=5/2$.  

\begin{figure*}
\includegraphics [angle=-90,width = 16cm]{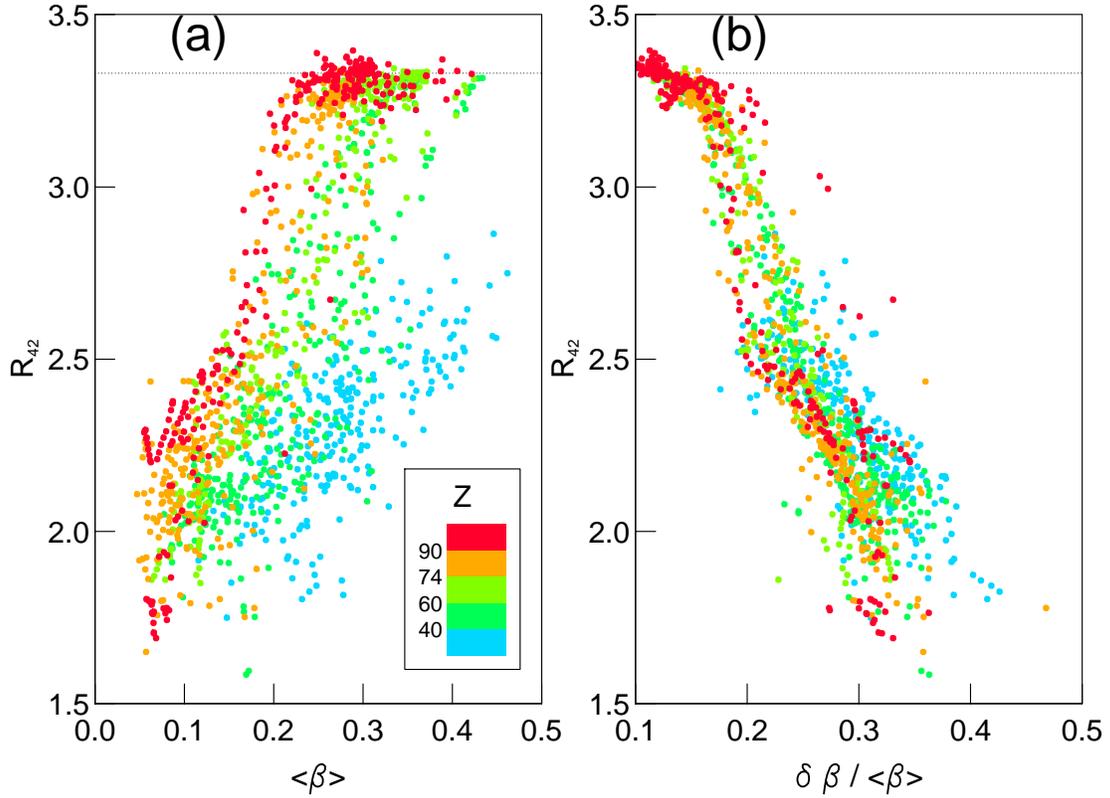}
\caption{(Color online) The ratio $R_{42}$ as a function of the mean deformation  
$\langle\beta\rangle$ (panel (a)) and the $\beta$-softness parameter
$\delta \beta /\langle \beta \rangle $ (panel (b)) for the calculated nuclei in their ground states.
Dotted lines show the rigid rotor value $R_{42} = 10/3$.
}
\label{fi:r42vdeltabeta}
\end{figure*}

It is interesting to see how well the physical structure indicator
$R_{42}$ correlates with the intrinsic shape properties of the
CHFB+5DCH wave functions.  Let us first examine the relationship
between $R_{42}$ and mean deformation $\langle \beta \rangle$.  This is 
shown in the left-hand panel of Fig.\ref{fi:r42vdeltabeta}.
One can see that the value of $\lb\beta\rb$ by itself does not determine
whether the yrast spectrum has a rotational character.  The $R_{42}$
has the rotational value for $\langle \beta \rangle$ in the range $0.2-0.45$, but
nuclei with nonrotational spectra are common with $\langle \beta \rangle$ values 
up to 0.3. In fact the highest value of $\langle \beta \rangle$ in our calculations
is found for a nucleus ($^{26}$Mg) for which $R_{42}=2.4$, both
theoretically and experimentally.  
Evidently,
what is needed as well to determine the rotational properties is a 
measure of the rigidity of the shape.  For that purpose, 
we use the $\beta$-softness parameter, $r_\beta =
\delta \beta /\langle \beta \rangle$.
The $R_{42}$ values are plotted with respect to $r_\beta$
in the right-hand panel of
Fig. \ref{fi:r42vdeltabeta}.  As may be seen from the figure, this
provides a much better separation between the rotational and nonrotational
spectra.  Effectively, the $\beta$-softness parameter should be less than
0.2 for a rotational spectrum.

\begin{figure}
\includegraphics [width = 7cm]{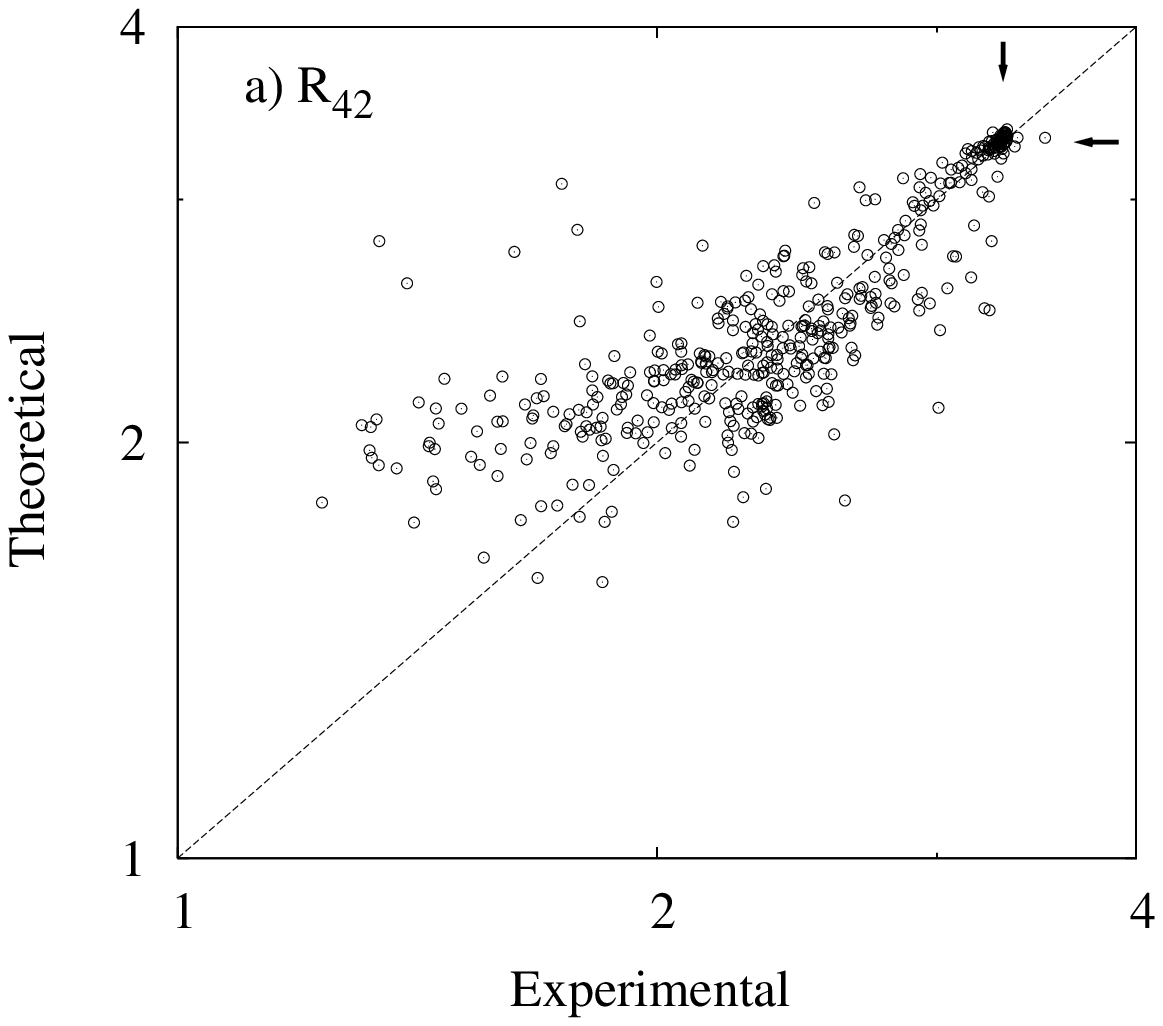}
\includegraphics [width = 9cm]{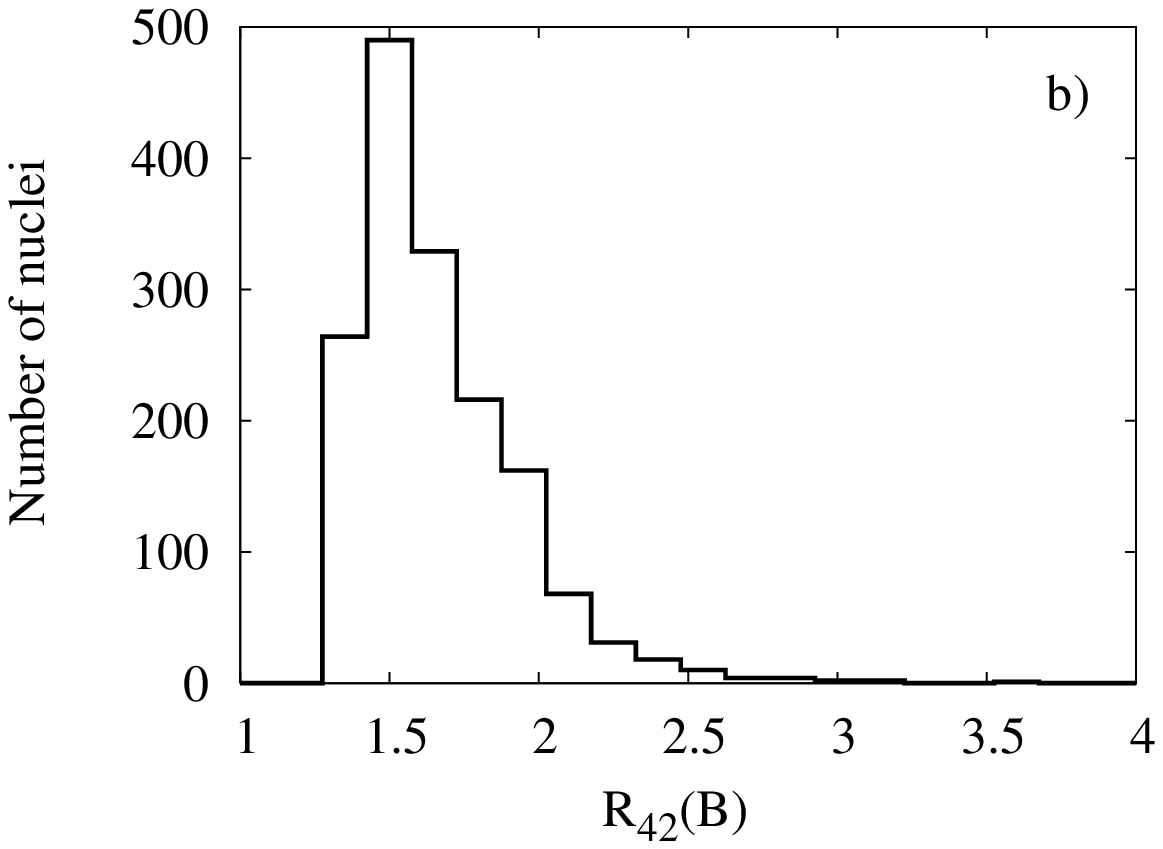}
\caption{Panel a): $R_{42}$ comparison of theory and experiment 
for 480 nuclei.  Arrows indicate the rigid rotor value $R_{42}=10/3$. 
Experimental data is from Ref.~\cite{brookhaven}. Panel b): distribution of
$R_{42}(B)$ for  CHFB+5DCH wave functions of \Nspectra nuclei.
}
\label{fi:r42_compare}
\end{figure}

We turn to the performance of the theory of the $4^+_1$ level, comparing
energies 
to experimental data.  Of the 484 nuclei
with tabulated experimental energies~\cite{brookhaven}, 480 meet the
criteria to be included in our theoretical data base. Left-hand panel of Fig.
\ref{fi:r42_compare} shows the comparison of the theory to experiment
as a scatter plot for $R_{42}$. 
For most nuclei the $R_{42}$ values in both measurements and calculations fall between 
$R_{42}$ = 2 and $R_{42}$ = 10/3 limits of the vibrational and rotational models, respectively.
Values of $R_{42}$ less than one are certainly possible when the spectrum
is dominated by two-quasiparticle excitations, which is common near
magic numbers.   
Statistical performance for the data set is given in Table \ref{ta:metrics}.  The average value of the 
ratio $R_{42}$ comes out very well, only 3\% higher than the
measured average.  The dispersion about the mean is also quite
good, better than the predictions for the absolute energies of
the $2^+_1$ excitations.

In the subsections below, we will analyze the properties of other
excitations with respect to the rotational character  of the ground
state.  Since the $R_{42}$ measure is very clear, we shall make much
use of it to examine the connections.

It is of interest to examine the transition strengths
$B(E2; 4^+_1\rightarrow 2^+_1)$, even in the absence of a critical
review and evaluated tabulation of the experimental data.
To interpret this quantity we
take the ratio to the $2^+_1\rightarrow 0^+_1$ transition,
defining $R_{42}(B)$ as
$$
R_{42}(B) = \frac{B(E2; 4^+_1\rightarrow 2^+_1)}{B(E2; 2^+_1\rightarrow 0^+_1)}.
$$  
Two anchor points to interpret $R_{42}(B)$
are the axial rotor model for which $R_{42}(B)= 10/7$, and the
harmonic vibrator model for which $R_{42}(B)= 2$.  The distribution of
calculated values is shown as a histogram in the right-hand panel of Fig. \ref{fi:r42_compare}. There is a peak at the
axial rotor value, but no peak at the vibrator value or
anywhere else.  The calculated $R_{42}(B)$ range from 1.43 for the nucleus $^{240}$Cm to
5.7 for the nucleus $^{180}$Pb.  There are no calculated nuclei with $R_{42}(B)$ 
smaller than the axial rotor value.

\subsection{The first $6^+$ excitation}

\begin{figure*}
\includegraphics [width = 12cm, angle = -90]{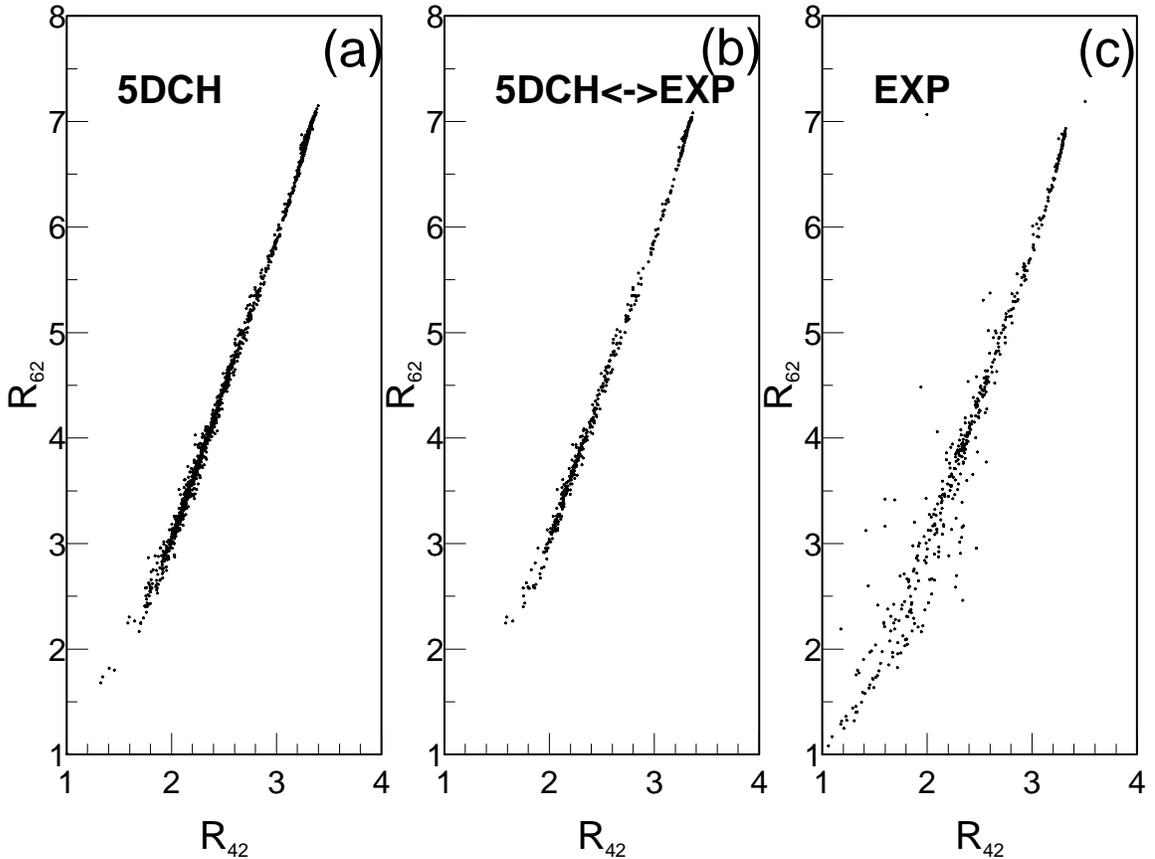}
\caption{Ratios of yrast excitation energies, $R_{62}$, as a function of
$R_{42}$.  Panel (a): CHFB+5DCH theory for \Ngoods nuclei. 
Panel (c):
experimental ratios for 456 nuclei, with data from Ref.~\cite{brookhaven}.
Panel (b):  theoretical values for the nuclei shown in the right-hand
panel.
}
\label{fi:r62}
\end{figure*}

  The last excitation we shall examine in the yrast spectrum is the $6^+_1$ level.
If the $0^+_1,2^+_1$, and $4^+_1$ levels form a band with energies 
close to the axial rotor limit, the $6^+_1$ state is also part of the band 
in the vast majority of cases.  Deviations of its energy
from the rotational limit can also be extrapolated from the $R_{42}$
values using the Mallman systematics \cite{ma59,bu08}, namely
the empirical correlation of the ratios 
$R_{62}=E(6^+_1)/E(2^+_1)$ and $R_{42}$.
The correlation associated with the CHFB+5DCH energies is shown in Fig.
\ref{fi:r62}, left-hand panel, based on theoretical energies from \Ngoods~ nuclei.  
The scatter plot follows a line from about $(R_{42},R_{62})=(1.5,2.0)$ 
to the value $(10/3,7)$ corresponding to the rotational limit.  
The plot shows an accumulation of points at the axial rotor limit, as well
as a somewhat broader peaking near $(2.3,3.7)$.  For orientation, the positions
of the $\gamma$-unstable limit and the harmonic vibrator limit are for $(5/2,4.5)$ 
and $(2,2)$, respectively.  The experimental scatter plot of $R_{62}$ vs.
$R_{42}$ is shown in the right-hand panel of
Fig. \ref{fi:r62}. It shows data
for 458 nuclei, obtained from the Brookhaven database \cite{brookhaven}.
We also show in the middle panel the calculated nuclei corresponding to the 
experimentally known ones. 
The experimental points form a line very much like the one seen in the theory
plot.  The correlation is also very narrow for the upper half
of the line, but it becomes broader at lower values of $R_{42}$ and $R_{62}$.  The
experimental plot extends to lower values than we find in the
theory.  One possible explanation is the neglect of two-quasiparticle
configurations in the theoretical wave functions.  Such configurations
can produce high angular momentum at relatively little energy cost,
and therefore can give values of $R_{42}$ and $R_{62}$ close to 1. 
%\footnote{That $R_{42}$ $\simeq$ $R_{62}$ is a typical feature of excitation spectra in the seniority scheme
%for single-closed-shell nuclei, that are never predicted in our model.}
Also, the nuclei
with such low $R_{62}$ values may 
have failed our criteria to keep in the theoretical database.
The global figures of merit of the 
observable $R_{62}$ are reported in Table \ref{ta:metrics}.  The
reliability of the theory is quite high, although it does not do as 
well as for the lower 2$^+$ and 4$^+$ yrast excitations.

\section{Non-yrast excitations}
\label{nonyrast}
This section will examine in some detail the physical properties of $0^+_2$ and  $2^+_2$
excited states
but is also concerned with their possible role as head levels of collective bands, traditionally referred to as 
$\beta$- and 
$\gamma$-vibrational bands, respectively.   In order to do this we will
also need to consider the $2^+_3$ excitation,  which can very often be
considered as part of a $K = 0$ excited band.  The specific indicators we will
examine in this context are the excitation energy with respect to
band head, the in-band transition rate compared to that for the ground state band, and the
relative out-of-band transition matrix elements.  Unfortunately, the
data tabulations do not exist to make a systematic comparison to experiment.
However, the band character has been much discussed in the rare-earth
region, and we can compare some out-of-band rates there.  
As the $2^+_3$ levels are almost systematically members 
of the excited K = 0 bands, an alternative to $\beta$-vibrational band 
interpretation
is suggested, namely that of coexisting band structure inside nuclei.  
A detailed discussion of the $\gamma$-degree of freedom and associated collective 
band excitations is deferred to a later publication.

\subsection{The $2^+_2$ excitation}

\begin{figure}
\includegraphics [width = 6cm, angle = -90]{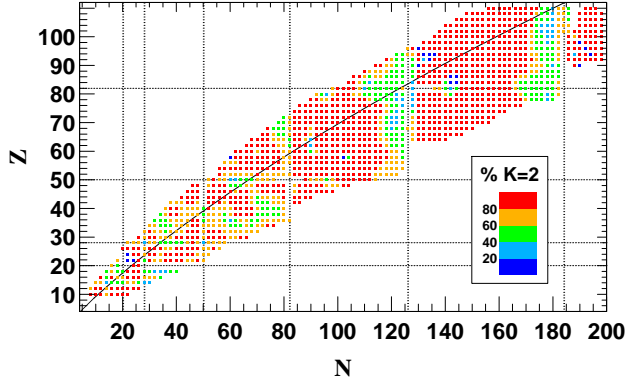}
\caption{(Color online) Chart of the computed nuclei showing the 
probability of $K = 2$ component in the wave function of the $2^+_2$ levels. The $2^+_2$
states with more than 75 $\%$ of $K = 2$ components in wave functions are
considered as $\gamma$-vibrations.  The black curve shows the beta-stability
line.
}
\label{fi:k2chart}
\end{figure}

\begin{figure}
\includegraphics [width = 8cm ]{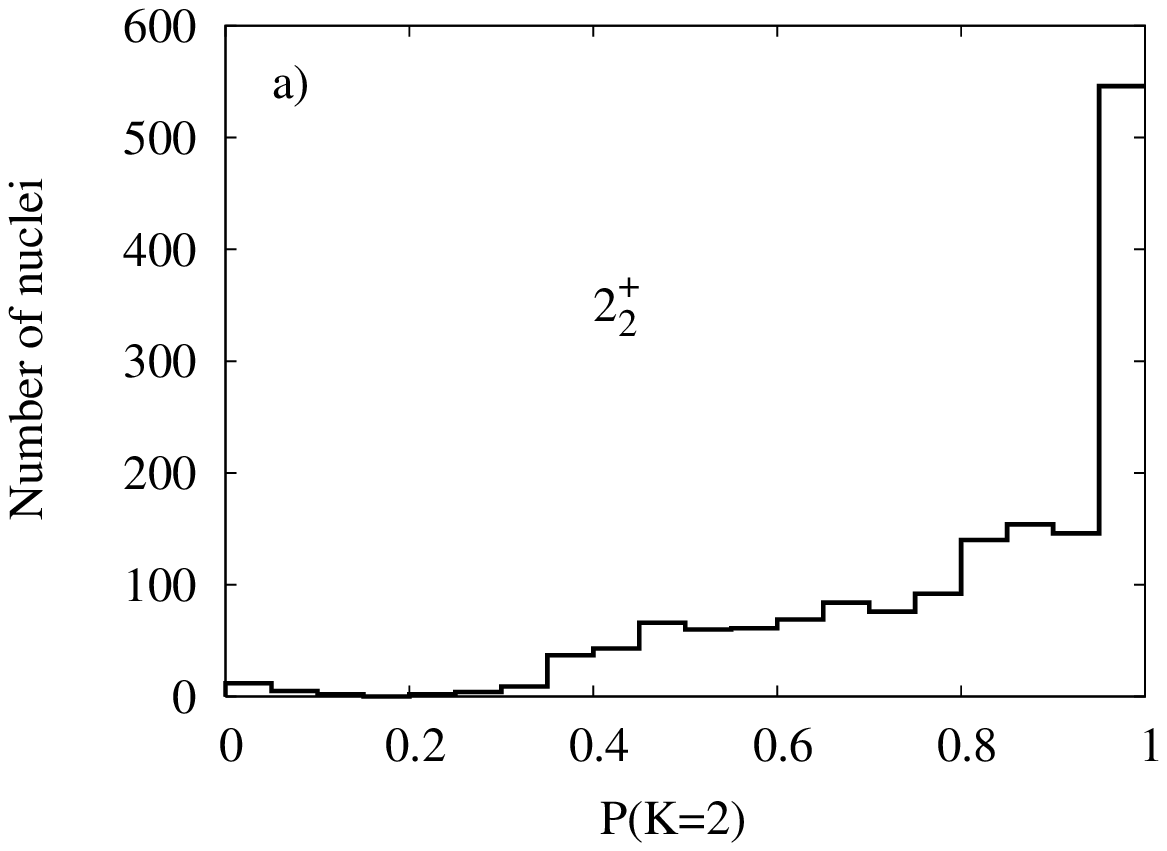}
\includegraphics [width = 8cm ]{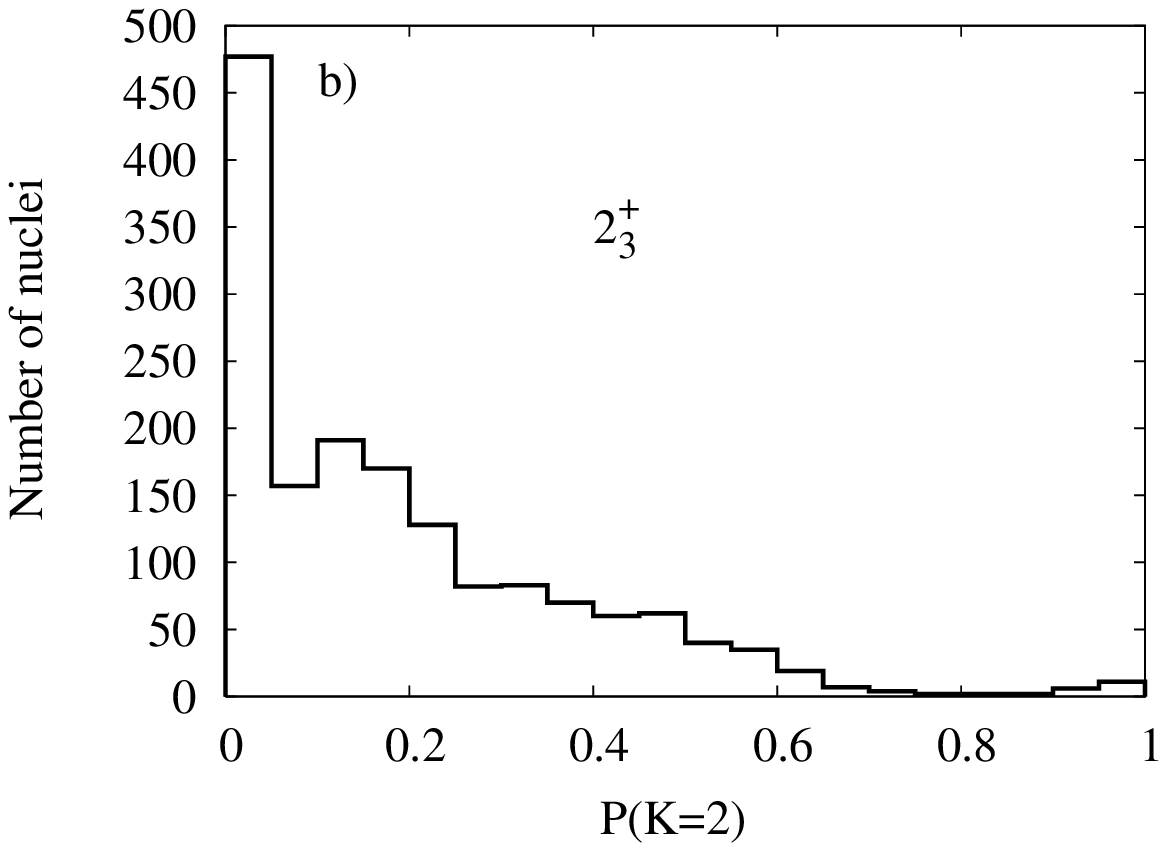}
\caption{\label{fi:K2}
Probability of $K = 2$ component in the wave functions of the second 
(panel a)
and third (panel b) excited $2^+$ states.
}

\end{figure}

The lowest non-yrast excitation typically
has $J=2$, and it is 
often interpreted in the collective model as a shape excitation
in the $\gamma$ degree  of freedom.  A theoretical indicator for
that character is the $K$-content of the wave function.  This is shown
visually in Fig. \ref{fi:k2chart} indicating the probability $P(K)$
by the coloring of the nuclides.  Apart from nuclei close to 
magic numbers,  the vast majority of second $2^+$ states have $P(K = 2)> 0.75$ 
and can be considered
as $\gamma$-vibrations. 
In the upper right corner of Fig. \ref{fi:k2chart} is a domain without coloring.
In this narrow mass region the inner potential barrier is not high enough to sustain excited
states, and the nuclei go to fission. 
For a more quantitative view of the $K$
distributions we show them by histograms in 
Fig.~\ref{fi:K2} for the second and third
$J=2$ states in the spectrum.
For the $2^+_2$ state (left-hand panel),
there is a sharp peak close to $P(K = 2)=1$, together 
with a broader distribution of lower probabilities.  
For the most part, the nuclei within the sharp peak have $R_{42}$ close to 
the axial rotor value.  Thus, for these nuclei we have a clear identification of the
$2^+_2$ level as a $\gamma$-excitation.  

The plot for the $2^+_3$ state in the right hand panel, shows that this level may be viewed as a $\beta$-excitation in many nuclei.  Here the strong peak is at $P(K = 2)=0$.   
Interestingly, there are a few nuclei for which the roles of the second 
and third state are reversed, as can be seen in Fig \ref{fi:k2chart}.  It happens that our example $^{152}$Sm in 
Sec. III is of this kind.  
In the discussion below, we will designate the second or third $2^+$ state with the
larger $P(K = 2)$ the $2^+_\gamma$ level, if $P(K = 2) > 0.75 $ for all nuclei with $R_{42} \geq 2.3$ even though we know that $\gamma-$vibration 
is a designation specific to well deformed nuclei.

\begin{figure*}
%\vglue -4 truecm
\includegraphics [width = 12cm, angle = -90 ]{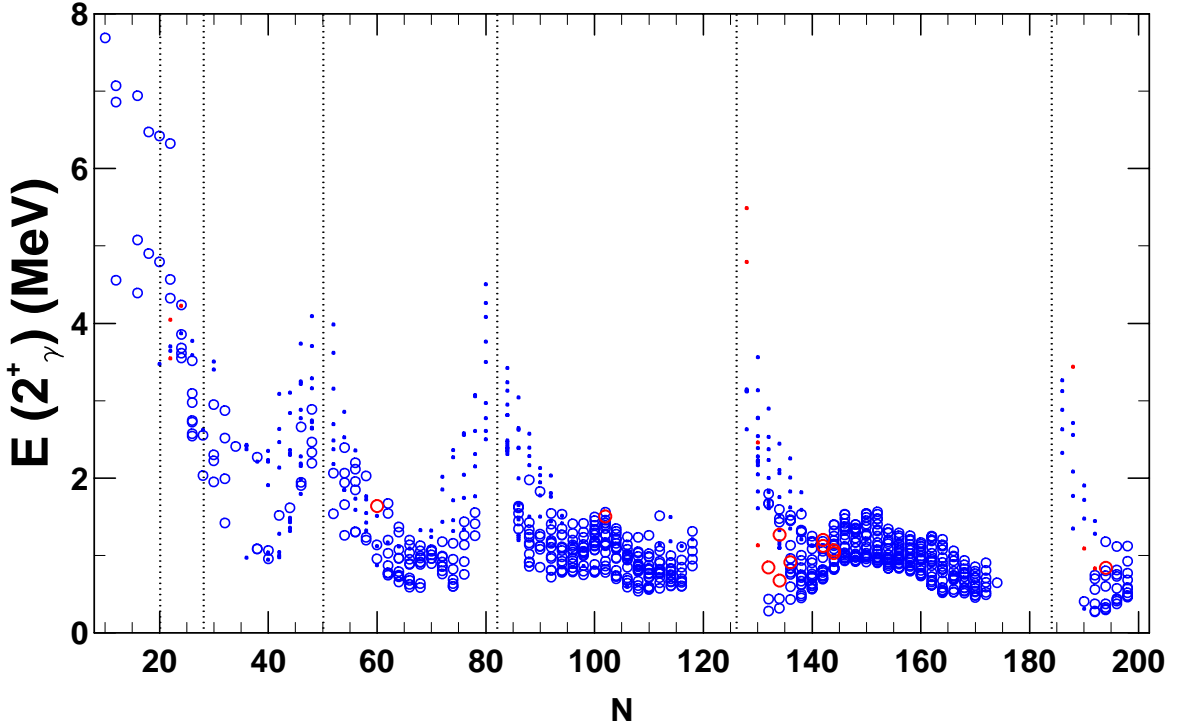}
%\vglue -4 truecm
\caption{\label{fi:egvN}
(Color online) Excitation energy 
of the CHFB+5DCH $2^+$ level with $P(K=2) \ge 0.75$ 
as a function of neutron number $N$. Open 
symbols in blue and red colors indicate the $2^+_2$ and the $2^+_3$ levels for nuclei with $R_{42} \ge 2.3$, respectively.
These levels are defined as $2^+_\gamma$ excitations. Dots are for nuclei with $R_{42} < 2.3$.
}
\end{figure*}

\begin{figure}
\includegraphics [width = 8cm , angle = -90 ]{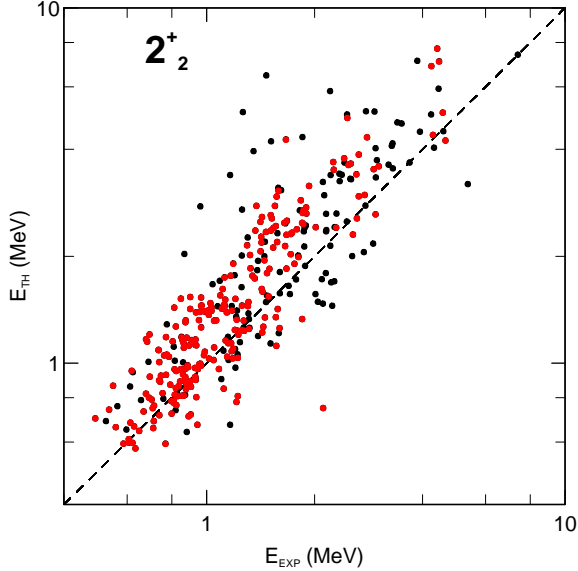}   
\caption{\label{fi:e22-c}
(Color online) Excitation energy of the second $J=2$ excitation, comparing 352 nuclei.
Experimental data is from Ref.~\cite{brookhaven}. The
$2^+_\gamma$ levels are marked with red color.
}
\end{figure}

The systematics of the 5DCH $2^+_\gamma$ energies 
are shown in Fig. \ref{fi:egvN} (open circles)
as a function of neutron number. The distribution of
excitation energies displays sharp structures with maxima near N$\geq$ 50 magic numbers, for which $R_{42} < 2.3$.  
Minima are found, as expected, half way between major closed shells
and they reach very low values ($E_x  \simeq 200$ keV) in heavy nuclei
with Z$ \geq$ 98.  
The performance of the CHFB+5DCH on the energies of the $2^+_2$ levels is shown in Fig. \ref{fi:e22-c} through comparing  
the calculations to the evaluated data for 354 nuclei~\cite{brookhaven}.   
The theory clearly reproduces the variation of the experimental energies,
which range 
over more than an order of magnitude.  
The colored symbols in this figure are for $2^+_2$ states identified as $2^+_{\gamma}$ levels
in the CHFB+5DCH calculations.
On average the theoretical 
energies are somewhat high.  The figures of merit, given in Table \ref{ta:metrics},
show that the theoretical energies average about 25\% higher than
the experimental ones.  Interestingly, the variance $\sigma$ is smaller than
that for the $2^+_1$ excitations.

\subsection{The $0^+_2$ excitation}
In the framework of the CHFB+5DCH theory, the $0^+_2$ excitation can arise in several
ways:  as a $\beta$-vibration, as a coexisting state of very different
shape, or something in between.   To be a $\beta$-vibration,
the excitation should have nearly the same $\lb\beta\rb$ as the ground state, but a larger
dispersion $\delta \beta$.  The excitation in this limit is very dependent on the
calculated mass parameters for the $\beta$ degree of freedom, which have
been calculated using the Inglis-Belyaev formula.  This treatment has
known deficiencies and we expect that the predicted excitation energies
would be somewhat lower if the Thouless-Valatin prescription were used.
The other likely structure for the $0^+_2$ excitation arises from the coexistence of vastly different
deformation at nearly the same energy.  The latter mechanism is prominent
in light doubly-magic nuclei~\cite{be03,be03a} and also in the actinides where the 
superdeformations occur at low excitation \cite{De06,kr92,ta98}.
A phenomenological signature of coexistence would be a low excitation
energy.  In fact there are a number of known nuclei for which the  $0^+_2$ level
is the first excited state, but we do not find such a low excitation
energy in our calculations.
However this observation does not at all mean that the present theory is not able to provide
reliable predictions for nuclei where shape coexistence and shape transition are present
and characterized by many measurements. 
Such features  are well described by our theory for the neutron-deficient Kr isotopes \cite{cl07,gir09}. 
That the present theory does not predict $0^+_2$ state as first excited state in a nucleus
obviously means that degrees of freedom other than collective quadrupole ones are at play and cannot be
ignored.

\begin{figure}
\includegraphics [width = 8cm]{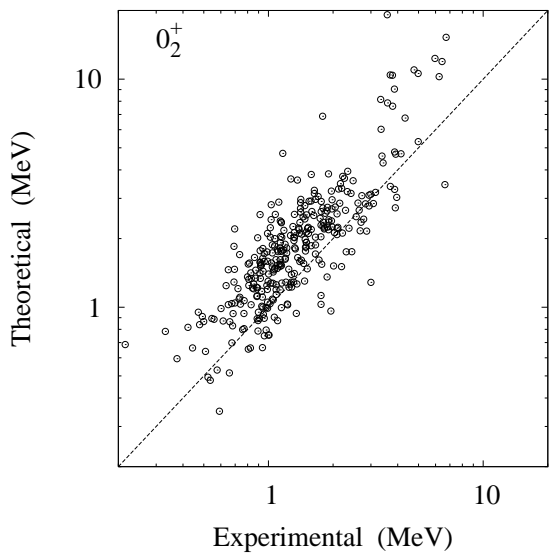}
\caption{\label{fi:e0}Excitation energy of the $0^+_2$ state compared with 
experiment \cite{brookhaven} .
}
\end{figure}

The theoretical and experimental
excitation energies of the $0^+_2$ excitation
are compared in Fig.~\ref{fi:e0}.  The experimental data 
set of 332 nuclei
was obtained from the Brookhaven data base \cite{brookhaven}.  Of the 332
tabulated nuclei, 317 are in the CHFB+5DCH calculated nuclei
and are shown in the Figure.
One sees that the theory reproduces the overall variation over one order of
magnitude, but that the calculations are systematically too
high. The figures of merit for the performance of the theory are given
in Table \ref{ta:metrics}.  The average $R_E$ is given by 
$\bar R_E=0.38$ and corresponds to predicted energies that are 
too high by $\sim50$\%.   The rms fluctuation about renormalized
theoretical energies is given by $\sigma_E$ in the table. Its value, 0.30,
corresponds to a fluctuation $+35\%-25\%$ in the error.   
\vskip 1cm

\begin{figure*}
\includegraphics [width = 12cm, angle = -90 ]{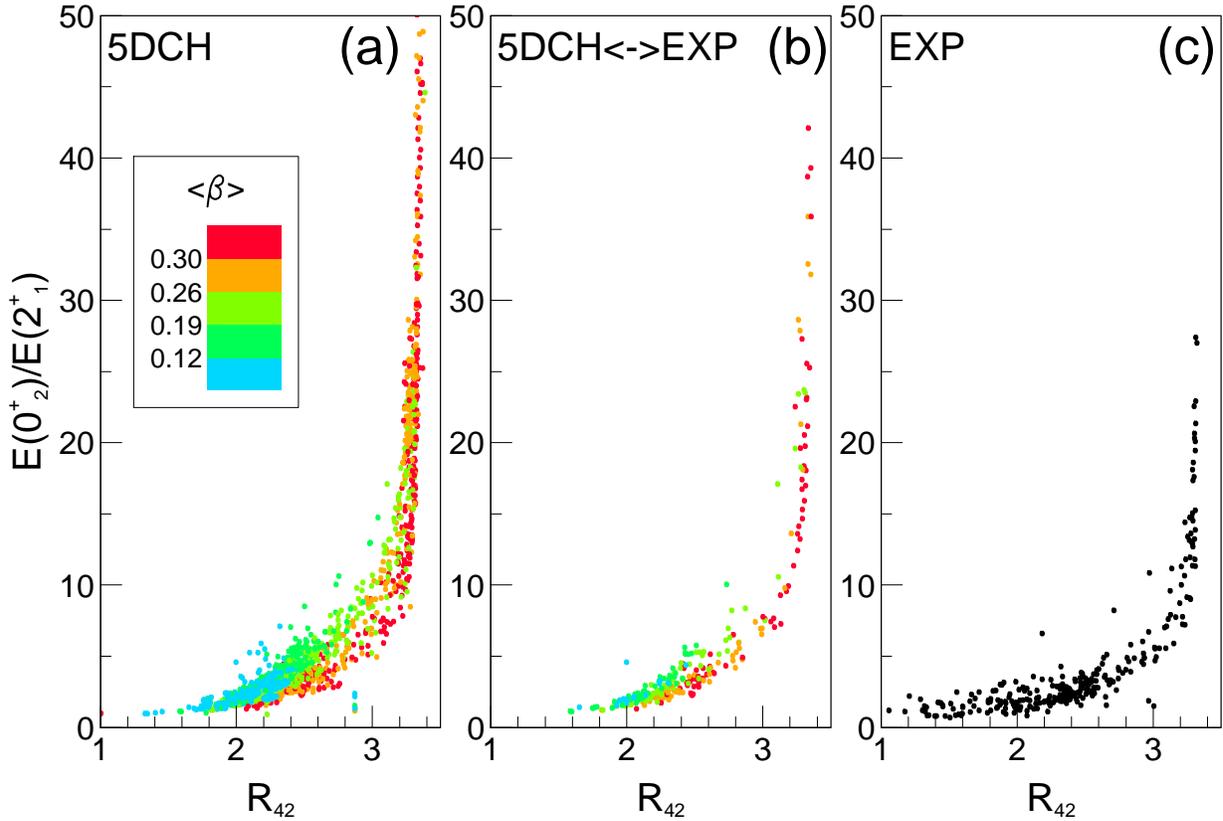}
\caption{\label{fi:e02v42}  (Color online) Panel (a): ratio $E(0^+_2)/E(2^+_1)$ as a function of
$R_{42}$ for the calculated nuclei. Panel (c): experimental data from Ref. \cite{brookhaven}.
Panel (b): CHFB+5DCH values for the nuclei shown in the right-hand panel.
Color code is for mean ground state deformation.
}
\end{figure*}

\begin{figure}
\includegraphics [width = 6cm, angle = -90 ]{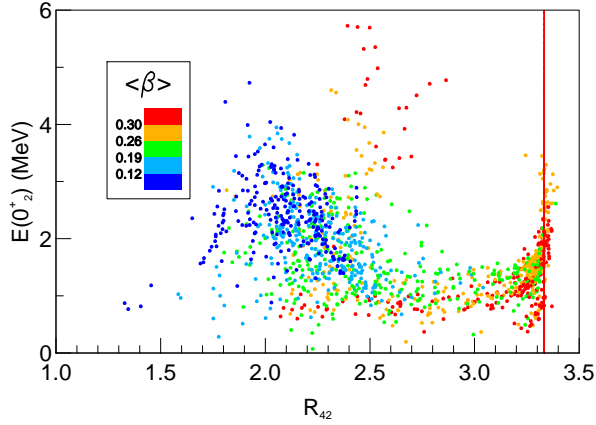}
\caption{\label{fi:e02vr42b}
(Color online) CHFB+5DCH excitation energy (MeV) of the $0^+_2$ state as a
function of the ratio $R_{42}$. Color code is for mean ground state deformation.
The vertical line (red color) indicates the rotational limit $R_{42}$ = $10/3$.
}
\end{figure}

We now examine the $0^+_2$ energy as a function
of deformation, following the work of 
Chou et al.
\cite{ch01b}.  These authors observed that there is a strong empirical correlation
between the ratio $R_{02}=E(0^+_2)/E(2^+_1)$ and $R_{42}$ measurements as shown in the right-hand panel of Fig.
\ref{fi:e02v42}. The left-hand panel shows a scatter plot of these quantities for the 
CHFB+5DCH calculations.  Indeed, one sees a strong correlation between the
two ratios. The $R_{02}$ ratio is flat with a value in the range 1-5
until $R_{42}$ approaches the axial rotor value, and then it can become
very large.  However, the width of the curve at fixed $R_{42}$ is too
broad to use this plot in a predictive way.  The overall shape of the
curve comes mostly from the variation of the denominator in the
ratio $E(0^+_2)/E(2^+_1)$ for both measurements and calculations.  We show in Fig. \ref{fi:e02vr42b} a
plot of $E(0^+_2)$ itself versus $R_{42}$, which is also informative.  
Here one sees three categories
of nuclei.  At one extreme are the spherical nuclei, having $R_{42}\sim 2$,
which have the highest excitations for the $0^+_2$ levels, typically in the
range of 2-4 MeV.  At the other extreme are the axial rotor nuclei with
$0^+_2$ energies in the range of 1-2 MeV.  Interestingly, the nuclei
in between favor lower energies, in the range of 0.5-1.5 MeV.  Those
nuclei are likely to be soft ones, and that would be reflected in both the
excitation energy of the $0^+_2$ levels and the range of values of $R_{42}$.
\subsubsection{Criteria for the occurrence of $\beta$-vibration}

As said above, the $0^+_2$ excitation in nuclei can arise as a $\beta$-vibration, 
as a coexisting level with deformation different from that for lower-lying states, or 
something intermediate between these two extreme structures. Here we would like to place the above discussion in a 
broader perspective. 

\begin{figure}
\includegraphics [width = 8cm]{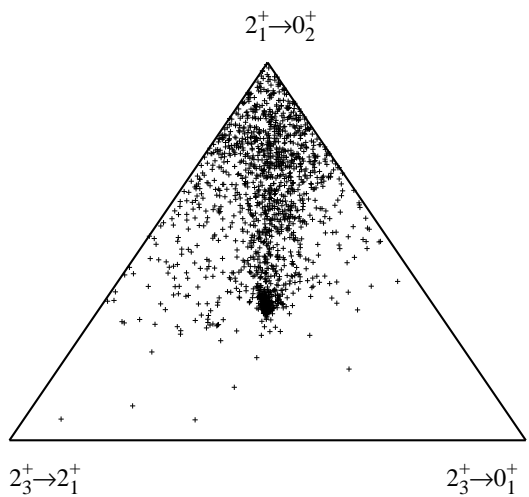}
\caption{Crossover matrix elements.  Relative magnitudes of the three
quantities 
$|M_{02}| , |M_{20}| ,  \sqrt{7\over 10 } |M_{22}|$ are shown by distances
to the sides of the triangle.  The vertexes
of the triangle correspond to the case where only the labeled transition
is nonzero.
}
\label{fi:triangle}
\end{figure}

\begin{figure}
\includegraphics [width = 6cm, angle = -90]{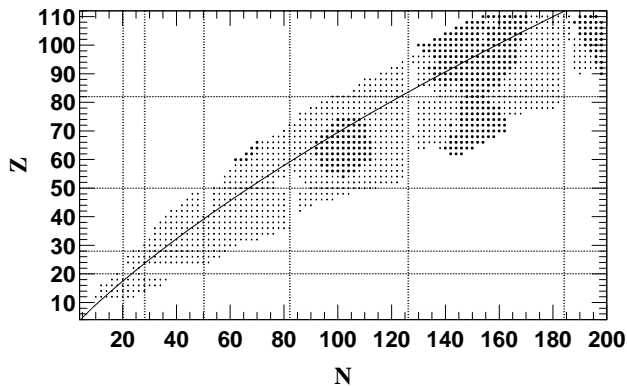}
\caption{Chart of nuclei (full circles) in the vicinity of the center point of the
triangle shown in Fig. \ref{fi:triangle}. The continuous curve is for the $\beta$-stability line, 
and dots are for nuclei between driplines as shown in the left-hand panel 
of  Fig. \ref{fi:chart1}.
}
\label{fi:jpd}
\end{figure}

First we consider relationships between quadrupole
transition matrix elements calculated in our theory for the $2^+_1 \rightarrow 0^+_2$, 
$2^+_3 \rightarrow 2^+_1$, and $2^+_3 \rightarrow 0^+_1$ transitions.
If the spectrum truly exhibits a $\beta$-vibrational band, the quadrupole
transitions between it and the ground state should be governed by
a single parameter, the matrix element of the quadrupole operator
between the two intrinsic states.  Under these circumstances the
spectroscopic transition matrix elements are related by  Clebsch-Gordan
coefficients, cf.~\cite[Eq. 4-219]{BM}:
\be
\langle\beta J_\beta ||{\cal M}(E2)||g J_g\rangle
= (2 J_g +1 )^{1/2} ( J_g 0 2 0 | J_\beta 0 ) \langle \beta | {\cal M}(E2)
| g\rangle.
\label{eq:gb}
\ee
Here the ground-state and the $\beta$-vibration bands are labeled by
$g$ and $\beta$, respectively.   We examine now the three CHFB+5DCH cross-band
transitions,  $(J_\beta,J_g) = (0,2),(2,0),(2,2)$, to see how well 
Eq.~(\ref{eq:gb}) is satisfied.  According to the model, the 
magnitudes $|M_{J_\beta,J_g}|$ should satisfy
\be
|M_{02}| = |M_{20}| =  \sqrt{7\over 10 } |M_{22}|. 
\label{eq:condition}
\ee
To display the deviations of the computed matrix elements from these
conditions, we take the ratio of the three quantities
$|M_{02}| , |M_{20}| ,  \sqrt{7\over 10 } |M_{22}|$
to their  total.  The fractions are plotted in 
Fig.~\ref{fi:triangle} as points
within a triangle, the fraction given by the distance to a side of the
triangle.  We see that there is a concentration
of points at the center point of the triangle; of the 1707 calculated
nuclei, 398 have values of the relative matrix elements within 15\%
of equality.  The distribution of these nuclei in $Z$ and $N$ is shown
in Fig.\ref{fi:jpd}.  One sees four regions where the condition is
well satisfied, including the strongly deformed rare earths and 
actinides.  
We conclude that the CHFB+5DCH theory predicts that 
$\beta$-vibrational bands should be quite common, taking as a 
criterion that eq.(26) be approximately satisfied.
There is also a
concentration of points at the upper apex of the triangle in Fig \ref{fi:triangle}. For these
nuclei, the $0^+_2\rightarrow2^+_1$ matrix element is much larger than
the two matrix elements involving the $2^+_3$ excitation, suggesting that
the excitations behave more like independent phonons.

\subsubsection{$0^+_2 \rightarrow 2^+_1$ transition}

We now focus on the experimental situation with respect
to the $0^+_2 \rightarrow 2^+_1$ transition strength.  This is an important
observable in an ongoing controversy about the existence of $\beta$-bands
in deformed nuclei \cite{Ga01,Wi08}.

In \cite{Ga01}, it is concluded
that the observed $\beta$- to ground state -band transitions are orders of magnitude
weaker than those predicted by collective models for deformed rare-earth
nuclei, except for very few.  To assess the performance of the CHFB+5DCH
at least in this limited region,
we have compared the CHFB+5DCH calculations of the 
$B(E2;0^+_2 \rightarrow 2^+_1)$ strengths with the experimental data on
the 9 nuclei compiled in Ref. \cite{Ga01}.  For all of these nuclei,
the CHFB+5DCH theory predicts a $\beta$-vibrational band that satisfies
the criteria discussed in the previous subsection.  For 4 of the nuclei 
($^{152, 154}$Sm, $^{154}$Gd,  and $^{168}$Yb),
the calculated B(E2)'s are of the same order as the
experimental ones, but somewhat higher by up to a factor of 2 or so.  However,
for the
remaining 5 nuclei ($^{158}$Gd, $^{166,168}$Er, 
 and $^{172,174}$Yb), the experimental values are an order of magnitude
smaller and in strong disagreement with the CHFB+5DCH theory.  Thus, for these nuclei
at least, the observed band built on the $0^+_2$ states does not
correspond to 
$\beta$-vibration calculated in the CHFB+5DCH theory.

Recent measurements have shown that many $0^+$ excited states are present
at low excitation energy in the deformed rare earths \cite{me06}.  This
suggests that the $0^+$ levels described by the CHFB+5DCH may
be quite fragmented.   For example, 
the $\beta$-vibrational mode couples to such modes as pairing vibrations 
and/or incoherent 2qp
excitations.  In the algebraic models, efforts have been made 
to explain the extra $0^+$ states by introducing many-particle 
many-hole excitations \cite{or99,he08}. Other regions of deformed nuclei like the actinides and
transactinides would be worth investigating to check whether they also
are missing the coherent $\beta$-vibrational structure predicted by 
the CHFB+5DCH theory.

\subsubsection{$0^+_2 \rightarrow 0^+_1$ transition}

\begin{figure}
\includegraphics [width = 12cm, angle=-90]{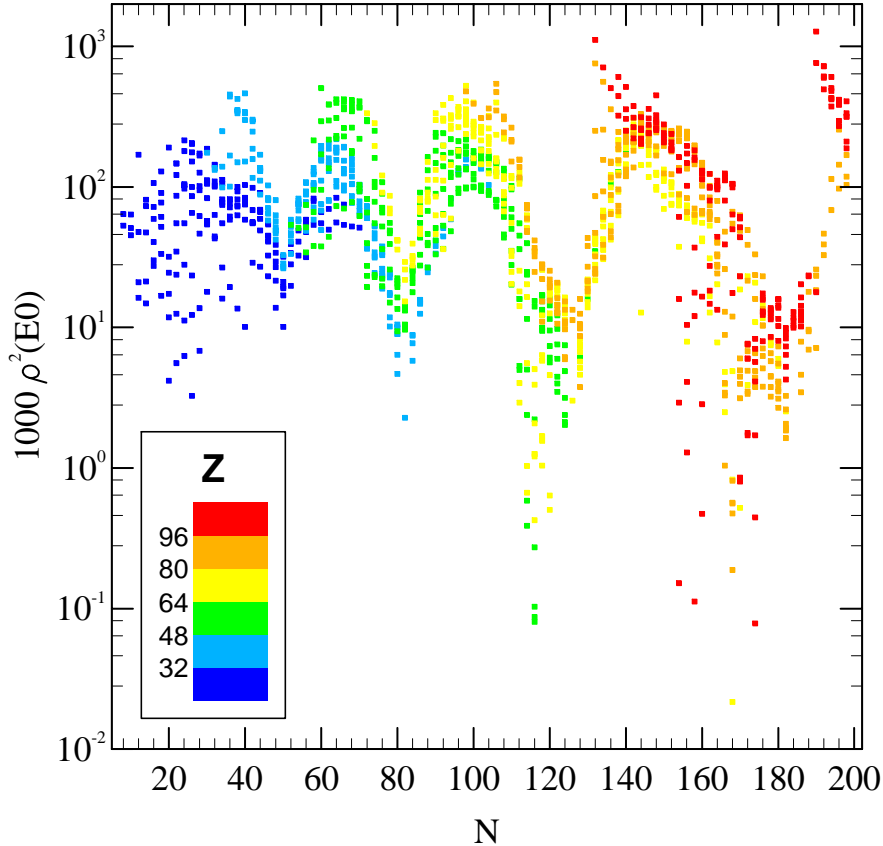}
\caption{\label{fi:E0vN} (Color online) Systematics of the calculated squared monopole transition matrix
element $\rho^2(E_0;0^+_2 \rightarrow 0^+_1) $ as a function of neutron number $N$.
Color code is for proton  number.
}
\end{figure}

Another observable relevant to the structure of the $0^+_2$ level is its monopole transition strength to the ground
state. The strength is conventionally expressed in terms of the
quantity $\rho^2$ defined as \cite{Wo99}
\begin{equation}
\displaystyle \rho^2(E0;0^+_2 \rightarrow 0^+_1)=|
\frac{\langle 0^+_2 |\sum_{i=1}^Z r_i^2| 0^+_1\rangle }{R_0^2}| ^2 ,
\end{equation}
with $R_0=1.2 A^{1/3}$ fm.  We calculate the required matrix element
 as
\be
\langle 0^+_2 |\sum_{i=1}^Z r_i^2| 0^+_1\rangle=
 \int d a_0 d a_2 g^{0_1}_0(a_0,a_2)g^{0_2}_0(a_0,a_2)\langle\Phi(a_0,a_2)
| \sum_{i=1}^Z r_i^2 | \Phi(a_0,a_2)\rangle .
\ee
We find rather interesting systematics with respect to the neutron number as shown in Fig. \ref{fi:E0vN}.
The calculations display oscillatory structures with broad maxima located near mid-shell closures 
(N~$\simeq$~40, 64, 100, and 150) and sharp minima in the vicinity of shell closures with N~$\simeq$~20, 28, 50, 82, 126,
and 184. Except for light nuclei, all these minima take place for mean ground state deformations $<\beta >$ with small values, that is for 
spherical equilibrium shapes. These features are globally consistent with IBA calculations for E0 transitions: the
$\rho^2(E0;0^+_2 \rightarrow 0^+_1)$ values raise sharply in the shape transition regions and then remain 
large for well deformed nuclei  \cite{Br04}. Other minima in E0 strengths are found below N $\simeq$ 126 and N $\simeq$ 184 neutron shell closures.
%in $\rho^2(E0;0^+_2 \rightarrow 0^+_1)$ values display fragmentations, thus suggesting that neutron shell closure
%in heavy nuclei might depend on proton number.
% This observation is in keeping with earlier HFB
%predictions \cite{Be01} for the superheavy mass region (Z $>$ 110), which indicate that the spherical Z = 114, 120, and 126
%shell gaps depend upon neutron number. The $\rho^2(E0)$'s calculated for N $>$ 184 and Z $>$ 100 
%display a backbend trajectory that is not yet
%understood.
Those associated with $N \simeq 162$ and weak $<\beta>$ values are correlated with the opening of the neutron shell closure in 
transactinide nuclei, as discussed previously in Sec. \ref{corener}.
The other two minima take place in open shell nuclei at mean ground state deformation in the ranges $< \beta > \ \simeq $ 0.19-0.26
and $< \beta > \ \simeq $ 0.26-0.30 for neutron numbers $N \simeq 116$ and 
$N \simeq 158$, respectively.
These features as well as similar ones identified in light nuclei with N $\simeq$ 20 and N $\simeq$ 28 indicate that the existence of minima
in the E0 strengths over the (N,Z) plane are not exclusively correlated with shell closure and spherical ground states.
Finally we note that the E0 strength values are at a maximum near N = 132 and N = 190 and decrease with N increasing.
These features are relevant to the Z $>$ 96 isotopic chains for which mean ground state deformations display strong variations 
(see right-hand panel in Fig. \ref{fi:chart1}).

We have compared
our $\rho^2(E0)$ transition strengths to 87 of the 91 nuclei tabulated
in Ref. \cite{ki05}.  The result for $R$ figure of merit is given in Table 
\ref{ta:metrics}.  We see that experimental matrix elements are on 
the average very small compared to theory.  This suggests that the
experimental $0^+_2$ levels may have a very different structure than the
calculated ones.  It may be that configurations ignored by the CHFB+5DCH, such as 2 qp excitations, may be
important in the non-yrast spectrum \cite{Ta83}. This problem with the parameter-free CHFB+5DCH theory which globally 
overestimates the E0 strengths by an order of magnitude is shared by other models of nuclear structure.
For example, enforcing realistic model descriptions of M1 and E2 transitions or charge radii, it is not uncommon that calculated
E0 strengths are up to ten times stronger than experimental values \cite{Ze08,Bo09,Ra09}.
The actual nature of E0 transitions remains an elusive issue pointing to major improvements
required in structure models.

\subsubsection{Coexistence between bands}

\begin{figure*}
\includegraphics [width = 12cm, angle = -90]{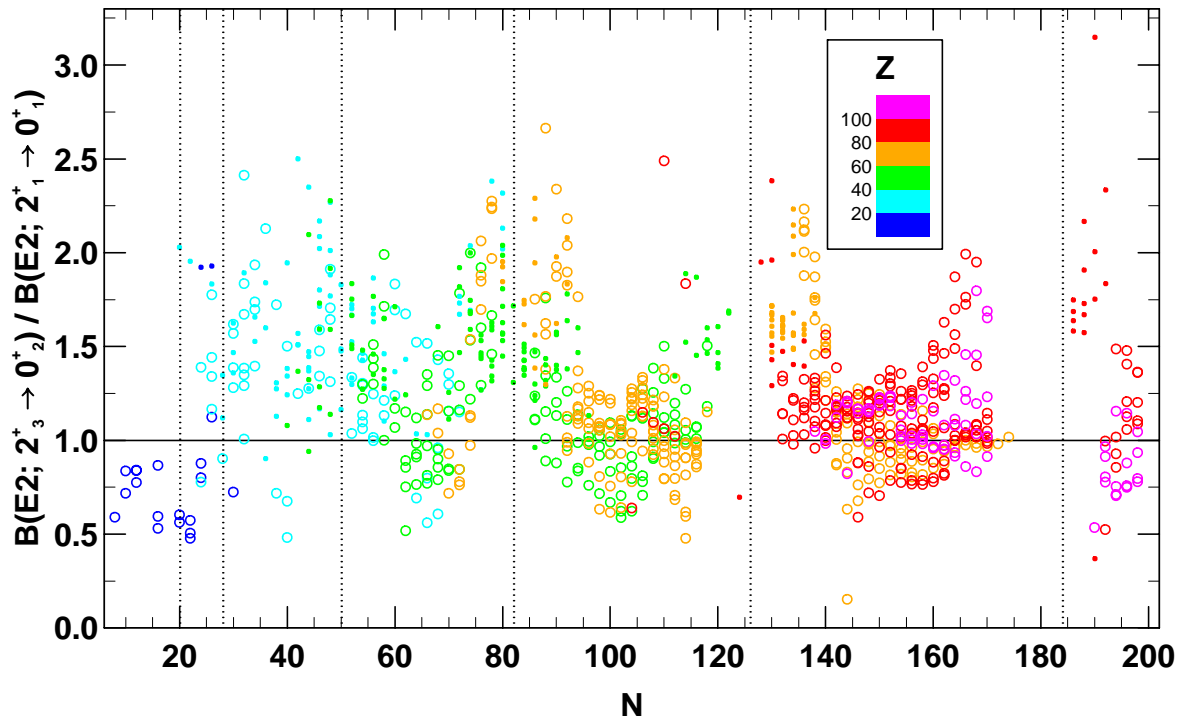} 
\caption{\label{fi:be23-NZ}
(Color online) The ratio $R_{20}(BB')$, Eq. (\ref{eq:rbb}), as a function of neutron number.
Color code is for proton number. Data marked with open and solid symbols are for $R_{42} \ge 2.3$ and $R_{42} < 2.3$, 
respectively.}
\end{figure*}

We have seen above that the conditions for the occurrence of $\beta$-vibration impose the medium- and
heavy-mass deformed nuclei to lay in specific 
(Z,N) regions (see Fig. \ref{fi:jpd}). One is then left with the issue as to what can be learned on 
collective excited K = 0 band properties for nuclei which do not belong to this sample. For this purpose
we define an important indicator of band structure through the ratio

\begin{equation}
\displaystyle R_{20}(BB')=\frac{B(E2;2^+_3 \rightarrow 0^+_2)}{B(E2;2^+_1 \rightarrow 0^+_1)}
\label{eq:rbb}
\end{equation}
for all nuclei of present interest with the provision that $2^+_3$ levels have
preponderant K = 0 component (i.e. P(K = 0) $>$ 0.75) in their wave functions which unambiguously makes them members of excited K = 0 bands. 
This ratio displays marked structures only if plotted versus neutron number. 
It is shown in Fig. \ref{fi:be23-NZ} where open and solid symbols are for $R_{42} \ge 2.3$ and $R_{42} < 2.3$, respectively.
$R_{20}(BB')$ values in the vicinity of $R_{20}(BB')$ =1  are representative of the points
concentrated at the center point of the triangle shown previously in Fig. \ref{fi:triangle}, and these 
take place near mid-neutron-closed shell numbers $N=100, 150$.
Most $R_{20}(BB')$'s take on values away from unity. Those with $R_{20}(BB')$ $>$ 1 are
suggestive of stronger collectivity present in excited K = 0 band than in ground state band, and the other way around 
for $R_{20}(BB')$ $<$~1 values.

It is for N $<$ 60 nuclei that the symbols in Fig. \ref{fi:be23-NZ} show disparate features as, within a narrow range
of N values, $R_{20}(BB')$ rapidly flips from $R_{20}(BB')$ $>$~1 to $R_{20}(BB')$ $<$~1. These features have not been analyzed in detail but suggest strong
shell effects driving nuclei from near spherical to deformed or from prolate to oblate, or vice versa. A typical
example is that offered by the neutron-deficient Kr isotopes
which display shape coexistence features, and also undergo a shape transition from prolate to oblate \cite{cl07}.
Nuclides with N $>$ 60 display less scattered features in their  ratios $R_{20}(BB')$ which now form a seemingly regular oscillatory trajectory
versus N. Extrema are not well localized but undoubtedly they are reasonably close to
$N= $ 66, 78, 90, 100, 116, 136, 146, 170 and 196 for nuclei with $R_{42} \ge 2.3$. Detailed information on
the location of nuclei with such properties over the (N,Z) plane is not yet available. At this moment we have checked that the ratio 
$R_{20}(BB')$ can serve as good indicator for the identification of nuclei known to display shape coexistence
(i.e. isotopes of the Se, Kr, Sr, Zr, Sm, Hg, Pb, and Po elements) associated with
the presence of two (or three) minima in potential energy surfaces. Coexistence between 
collective bands are calculated for Pd, Cd, and Te isotopes 
which are known experimentally to display such features, see e.g. \cite{lh99}. For these nuclides, coexistence is related to well localized maxima 
in collective masses present at different loci over the $(\beta$,$\gamma$) plane and not to prominent minima in potential energy
surfaces.  Other theoretical treatments of this kind of coexistence have
mostly been based on the Interacting Boson Model, invoking particle-hole
excitations between shells to explain the intruder states 
~\cite{ri87,Co96,Le97}.

\section{Summary and Outlook}
\label{summary}
    In this work we have evaluated the performance of the CHFB+5DCH
theory based on the Gogny D1S interaction as a global theory of nuclear 
structure. Highlights of the successes of the theory are its
accurate predictions with respect to charge radii,
the classification of nuclei as deformed rotors or not,
and the ground state band properties of the axial rotors.

The calculated 2-nucleon separation energies are interesting in that
they show shell effects and how they are modified for nuclei far
from stability.  An example is the nuclei near $N,Z=40$.  These are
often predicted to be spherical in HFB.  This is also the case for
the Gogny interaction, but we find that there is a change in structure
going to the CHFB+5DCH extension, and the nuclei become deformed rotors,
in agreement with experiment. The CHFB+5DCH theory is a suitable framework 
where the issues of shell erosion and shell collapse can be addressed.

As a spectroscopic theory, the CHFB+5DCH has considerable predictive power
for the lowest yrast and yrare states, even for nonrotational spectra.  
This is illustrated by predicted energies of the $2^+_1,4^+_1$ and 
$2^+_2$ excitations, which we compared with compiled empirical data.
The performance with respect to the $0^+_2$ levels also showed predictive power,
but here systematic deficiencies of the theory become apparent.
All in all, it is quite remarkable that a theory based on a many-body 
Hamiltonian with only 14 interaction parameters has such predictive power over
the broad range of the nuclei that can be treated in the methodology.
An important finding is that deformation
alone is not a good predictor of rotational spectra.  We defined
a quantity, $\beta$-softness, that correlates much better.
For convenience, we summarize in Table \ref{ta:metrics} the figures of merit for
the performance of the theory for excitation energies and transition
properties. We also provide in a retrievable form the specific predictions for spectral
properties of about 1700
even-even nuclei and analyzed some of the systematics of the 
predicted quantities.  
\begin{table}
\caption{\label{ta:metrics}Summary of the performance statistics of the CHFB+5DCH 
for excitation energies and transition properties.
See text for definitions of $\bar R$ and $\sigma$. The column ``Number"
gives the number of nuclei in the comparison data set.
}
\begin{tabular}{|l|c|cc|}
\colrule
Observable &   Number & $\bar R$ & $\sigma$ \\
\colrule
$E(2^+_1)$   &   513 & 0.11 & 0.35 \\
$B(E2;2^+_1\rightarrow 0^+_1)$ & 311  & 0.20  & 0.42  \\
$R_{42}$     &    480   &    0.03   &  0.14 \\
$R_{62}$     &     427    &  0.08  & 0.21 \\
$E(2^+_2)$ & 352 & 0.19 & 0.30  \\ % new
$E(0^+_2)$ & 317  & 0.31 & 0.36 \\%new
$ \langle 0^+_2| r_p^2 | 0^+_1\rangle$ & $87$  & 2.1 & 1.9 \\   % new
\colrule
\end{tabular}
\end{table}

There are a number of avenues that could be pursued to improve the
theory, some of which are quite straightforward, at least in principle.  
The treatment of the inertial masses could be improved by using the
Thouless-Valatin prescription which is better justified than
the cranking approximation we have used up to now.  This would surely
improve the energies of the $0^+_2$ excitations, which is one of the
problems of the current theory.  This would require calculating the
QRPA response function at every grid point.  This would of course
add to the computational burden, but the ingredients to perform the
calculation are available for the most part in the intermediate
calculations already performed.  The rotational inertias could also
be improved by using a finite amplitude rotational field $\omega \hat{J}_k$
adjusted to self-consistency for the angular momentum value being calculated
\cite{ma75}.  The ingredients of this self-consistent treatment
are already available and have been implemented in some calculations
\cite{Ob05}. This also would considerably add to the computational burden.

We have taken
the correlation energy as an indicator of the validity of the
Gaussian Overlap Approximation, and excluded nuclei whose calculated
correlation energies are unphysical.   A 
better treatment of the correlation energy would include the ZPE term coming
from the curvature of the potential energy surface.  This term is
small in nuclei with broader collective potential energy surfaces,
but in noncollective nuclei the minima can be narrow, and this
potential term in the Hamiltonian mapping might have a significant
effect.  

Another deficiency of the CHFB+5DCH that became evident in the discussion of
the $0^+_2$ level properties is the need for
2 qp components in the wave functions.  This can be carried out in 
the GCM if the Hamiltonian operator in the collective space is calculated
with full treatment of the nonlocality \cite{di76}.  However, 
there is no clear road to us for how to include these components and
keep the Gaussian Overlap Approximation as yet.

While it is remarkable that the Hamiltonian based on the Gogny D1S
interaction  has so much
predictive power after 30 years, one can still ask how the calculated
observables depend on the intrinsic properties of the Hamiltonian and
whether an improvement can be made at that level.   We note that there has already been some
work to improve the Gogny interaction for calculating nuclear masses, while keeping
unaltered the performance of the CHFB+5DCH and RPA theories achieved previously with D1S
\cite{hilaire}.

Finally, as a separate publication, we will present in
more detail the spectroscopic properties to higher angular momenta
for the deformed rare-earth nuclei.
This will also include discussion on the systematics of interband transitions and 
odd-even angular momentum staggering of the $\gamma$ bands~\cite{NewWork}.

\section*{Acknowledgment}
GFB thanks A. Bulgac, R. F. Casten, C. Johnson, L. Dieperink and L. Prochniak for 
discussions, and A. Sonzogni for help with the BNL database.
This work was supported in part by the UNEDF SciDAC Collaboration under DOE
grants DE-FC02-07ER41457 and DE-FG02-00ER41132. The authors are thankful 
to CEA-DAM Ile-de-France for access to CCRT supercomputers.


\begin{thebibliography}{100}
\bibitem{fix} 
	    For the Sr and Zr isotopic chains, we used the data of Refs. \cite{ca02,bu00}
	    to determine the experimental radii, taking the values of $^{88}$Sr
	    and $^{90}$Zr from Ref. \cite{an04}.
\end{thebibliography}
\end{document}